\documentclass[article, aps]{revtex4-1}
\usepackage{graphicx}

\usepackage{color} 
\usepackage{soul} 
\usepackage{amssymb}
\usepackage{amsmath}

\usepackage{bm}

\def\be{\begin{eqnarray}}
\def\ee{\end{eqnarray}}

\newcommand{\nn}{\nonumber\\}

\begin{document}

\title{ Transition Path Theory from Biased Simulations}
\author{G. Bartolucci$^{1}$, S. Orioli$^{1,2}$ and P. Faccioli$^{1,2}$}
\affiliation{$^{1}$ Physics Department of Trento University, Via Sommarive 14, 37123 Povo (Trento), Italy}
\affiliation{$^{2}$ INFN-TIFPA, Via Sommarive 14, 37123 Povo (Trento), Italy.}
\begin{abstract}
Transition Path Theory (TPT) provides a rigorous   framework to investigate the dynamics of rare thermally activated transitions. In this theory, a central role is played by the forward committor function  $q^+(x)$, which provides the ideal reaction coordinate. Furthermore, the reactive dynamics and kinetics are fully characterized in terms of two time-independent scalar and vector distributions.  
In this work, we develop a scheme which enables all these ingredients of TPT to be efficiently computed using the short  non-equilibrium   trajectories generated by means of a specific combination of enhanced path sampling techniques. In particular, first we further extend the recently introduced Self-Consistent Path Sampling (SCPS) algorithm in order to compute the committor $q^+(x)$. Next, we show how this result can be exploited in order to define efficient algorithms which enable us to  directly  sample the transition path ensemble.   
 \end{abstract}
\pacs{Valid PACS appear here}
\maketitle

\section{Introduction}
Biomolecules undergo thermally activated conformational transitions in order to reach their biologically active state or to perform functions \cite{book}. Understanding the physico-chemical mechanisms which control the kinetics of these processes is a central problem in the fields at the interface  between physics, chemistry and molecular biology. 
For most reactions of biophysical or biological interest,  however, identifying the reaction mechanism and estimating the reaction rate is  a  challenging task, both from the experimental~\cite{eaton} and the computational~\cite{anton, pande} standpoint.  The main reason is  that barrier-crossing processes are extremely  fast and rare events. For example, in  protein folding the average time it takes to complete a reactive event  is of the order of a few  microseconds~\cite{eaton}, while the  folding time can range from fractions of milliseconds to minutes or beyond.
 
Because of the high computational cost of performing plain MD simulations~\cite{anton}, alternative theoretical and computational frameworks are continuously being developed to efficiently characterize conformational reactions in complex and rugged energy landscapes (for a recent review  see e.g. \cite{elberrev}). An incomplete list of these techniques which are specific for reaction kinetics includes Markov State Models~\cite{MSM},  Milestoning~\cite{milestoning}, Transition Path Sampling \cite{TPS}, Transition Interface  Sampling \cite{TIS} and Forward Flux Sampling \cite{FFS}, along with different methods based on biasing the dynamics to promote reactive events \cite{steered, targeted, rMD1, BFA, MetaKin, rMD2}. 

In parallel with the advance of computational methods, theoretical frameworks need to be developed  in order to reduce the resulting data,  provide insight into the reaction mechanism and produce predictions for kinetic observables. 
In this context, Transition Path Theory (TPT)  \cite{TPT0, TPT1, TPT2}, briefly reviewed in appendix \ref{TPTsect}, displays several attractive features. For example,  in this theory \emph{time averages} are replaced with \emph{phase-space averages} defined over two stationary scalar and vector distributions: the transition path density  distribution $m_T(x)$ and the transition current  $J^i_T(x)$. At the same time, TPT also  rigorously extends and  exploits some of the key concepts of Transition State Theory \cite{TST1, TST2}, providing a rigorous  definition of the transition state which can be applied to rugged energy landscapes.

A pivotal concept in TPT is the so-called forward committor function $q^+(x)$; this collective variable  provides the ideal reaction coordinate and expresses the probability that a trajectory initiated at the point $x$ reaches the product state before returning to the reactant. It can be shown that both the transition path density distribution $m_T(x)$ and the transition current $J^i_T(x)$ can be formally expressed in terms of the forward committor function $q^+(x)$ and the Gibb's distribution, $\exp[-\beta U(x)]$. Therefore,  the  practical usefulness of TPT depends on the feasibility of accurately estimating the forward committor.

The Finite Temperature String Method~(FTSM) developed in Ref.s \cite{string2, string1,stringrevisited} provides a  framework to compute $q^+(x)$ by focusing on the so-called  principal curves. These one-dimensional manifolds identify the regions of configuration space which are explored by the transition pathways. In particular, for diffusion in smooth energy landscapes, the principal curves reduce to the minimum-free-energy paths (MFEPs)  \cite{string1, string2}. In the vicinity of these curves,   the iso-committor hyper-surfaces  are  identified with the $3N-1$ dimensional hyper-planes locally orthogonal to the nearby MFEP. The FTSM scheme sets the stage for performing practical calculations, and it is very valuable to investigate  transitions occurring in molecular systems. On the other hand, for conformational transitions as complex as protein folding, the application of the FTSM may be problematic, as the final results  might retain a  dependency  on the choice of the initial  guess for the principal curve.  This problem is also shared by other path sampling methods based on the numerical optimization or sampling of   some  functional  of the path (see e.g.  Ref.s \cite{DRP0, sdel, donniach, DRP1, DRP2, BFA, GD1, GD2, NEB, TPS}).

In this work, we develop  several theoretical and computational schemes to overcome some of these  limitations,  based on  combining  different procedures. The first result consists in showing how the SCPS approach introduced in Ref. \cite{SCPS} can be used to accurately estimate the committor function $q^+(x)$. This enhanced path sampling algorithm has  been already applied to the study of protein folding using an all-atom force field. Thus, the possibility of computing $q^+(x)$ from SCPS can bring valuable insight into  a number of complex transitions. 

Next, we discuss how the knowledge of the committor can be  exploited in order to increase the sampling efficiency of the transition path ensemble. To this goal, we use the Green's function formalism to re-derive a modified Langevin equation --originally introduced in Ref.  \cite{CLD0} --  which yields an exact sampling of $m_T(x)$. However, when the reaction mechanism is complex, sampling this distribution may still be  computationally very challenging. To cope with this problem, we define a  special kind of ratchet-and-pawl  Molecular Dynamics (rMD) \cite{rMD1, rMD2}, with a \emph{history-dependent}  biasing force defined in terms of the committor function.  We show that, in the limit of strong bias force, this dynamics can be used to compute many short reactive trajectories that travel only forward along the committor and sample the Boltzmann distribution restricted to the reactive region.

The paper is organized as follows. In section \ref{SCRsect} we introduce a specific range of time scales which we refer  to as the Steady Current  Regime (SCR).  We show that all TPT results can be accurately estimated from statistical averages performed over many short barrier-crossing trajectories,  with duration  in this time regime. This result is the key to make contact with the enhanced path sampling methods we use to estimate the committor $q^+(x)$ and to sample the transition path density $m_T(x)$. 
In section \ref{SCPSsect} we discuss how the SCPS algorithm can be used to estimate the committor, while in 
section \ref{CLDsect} we discuss  the algorithms to  sample the transition path density and the Boltzmann distribution, in the transition region. Section \ref{example} is devoted to illustrate and validate these results on a simple toy model,  while the main results and conclusions are summarized in section \ref{conclusions}.

\section{TPT from Non-Ergodic Trajectories}\label{SCRsect}

The  investigation of rare transitions occurring in complex systems is often computationally unfeasible without relying on enhanced sampling techniques. On the one hand, these methods are designed to efficiently produce \emph{short} reactive trajectories, without having to waste computational time in simulating thermal oscillations in the reactant state. On the other hand,  all the results of TPT are based on analyzing an infinitely long  (i.e. ergodic) trajectory. 
Thus, an important preliminary step towards efficiently computing $q^+(x)$,  $m_T(x)$ and $ J^i_T(x)$ consists in showing how these functions can be obtained from a statistical analysis of short (i.e. non-ergodic) reactive paths. As we shall see, this connection can be established only under specific assumptions about the relaxation time scales of the system. 

For sake of simplicity, in this work we specialize to systems obeying the overdamped Langevin equation 
\be\label{Lang}
\dot x= -\frac{D}{k_BT} \nabla U(x) + \eta(t),
\ee
where $x=({\bf x}_1, \ldots, {\bf x}_N)$ is a point in the $3N$-dimensional configuration space,  $U(x)$ is the molecular potential energy and $\eta(t)=({\bf \eta}_1(t), \ldots, {\bf \eta}_N(t)) $ is a memoryless white noise, obeying the fluctuation dissipation relationship $\langle \eta_i(t) \eta_j(0) \rangle = 6 D \delta_{i j}\delta(t)$. 
The probability density distribution $P(x,t)$ associated to Eq. \eqref{Lang} obeys the Fokker-Planck (FP) equation,
\be\label{SFP}
\frac{\partial}{\partial t} P(x,t)= \hat{H}_{FP} P(x,t),  
\ee
where 
$ \hat{H} _{FP} =  D \nabla \cdot \left(\nabla  + \beta \nabla U(x)\right) $ is  the FP  operator. 

The spectrum of  $\hat H_{FP}$  defines  the relaxation frequency scales of the system. In the presence of thermal activation, the spectrum is gapped; we shall  denote with  $k_F$  a typical fast frequency scale representing  the  eigenfrequencies above the gap. These modes  are associated with \emph{local} relaxation processes within metastable states (or within local networks of states). Similarly, we shall denote with $k_S$  the typical scale of the slow eigenfrequencies below  the gap, which are associated with  the global relaxation process. In particular, by focusing on the case of reactions with a single activation free energy barrier (two-state kinetics), we restrict our attention to spectra with a single non-vanishing soft mode $k_S$.

Since we are interested in the slow relaxation kinetics, we imagine to prepare a set of initial configurations in the reactant state, and  then to integrate Eq. (\ref{Lang}) starting from each of them.
After a time interval of the order of $t\sim k_F^{-1}$,  the  distribution of of initial configurations relaxes to a metastable  distribution,  which is locally Gibbsian and  confined within the reactant state:
\be\label{localGibbs}
P(x,t) \simeq  \rho_0(x) \equiv \frac{e^{-\beta U(x)}}{Z_R}~h_R(x), \qquad \left(Z_R = \int dx \frac{e^{-\beta U(x)}}{Z_R}~h_R(x)~\right).
\ee
For times  $t\gtrsim  k_F^{-1}$ a  probability current 
\be\label{JFP}
 J^i(x,t) = - D (\nabla^i + \beta \nabla^i U(x) ) P(x,t)
\ee
begins to flow from the reactant to the product, driven by spontaneous barrier-crossing events.  The current in Eq. \eqref{JFP} relaxes to $0$ only at very  long times,  $t\gtrsim k_S^{-1}$, when global thermal equilibrium is finally attained. This is the ergodic regime, which is invoked in the  original formulation of  TPT.

However, since the spectrum of the FP operator is gapped, it is possible to consider time intervals $\tau$ which are at the same time sufficiently long to ensure  local thermal relaxation within the metastable states,  yet sufficiently short  for the system to be still  far from the ergodic regime, i.e.
\be\label{SCR}
\frac{1}{k_F} \ll \tau \ll \frac{1}{k_S}.
\ee
Note that $k_S \tau \ll1 $ and $(k_F \tau )^{-1}\ll 1$ provide two independent  small expansion parameters, which will be exploited in the following in order to define systematic approximations. 

In the  kinematic regime (\ref{SCR}) the system has had time to perform \emph{at most} a single barrier crossing transition. At this time scale, the time-evolution of the probability current $ J^i(x,t)$ defined in Eq. (\ref{JFP}) is frozen.  To prove this,  let us  expand  $ J^i(x,t)$  in terms of the right eigenstates $R_n(x)$ of the FP equation:
\be
 \left.\frac{\partial}{\partial t}  J^i(x,t)\right|_{t=\tau} = \sum_n c_n ~e^{-k_n \tau} ~D( \nabla^i + \beta \nabla^i U(x) ) R_n(x), \qquad~(~
\hat H_{FP} R_n(x) = k_n R_n(x)~) 
\ee
If $\tau$ satisfies the inequality (\ref{SCR}), the contribution to this sum coming from all the  eigen-modes with eigen-frequency of order $\sim k_F$ is exponentially suppressed, since $k_F \tau \gg 1$. On the other hand, the single exponential factor containing the single slow eigenmode $k_S$ is $\simeq 1$, since $k_S \tau \ll 1$. Consequently, the probability current is approximatively stationary:
\be
\tau \left.~ \frac{\partial}{\partial t}  J^i(x,t)\right|_{t=\tau} \simeq 0.  
\ee
For this reason, the time interval (\ref{SCR}) will be referred to as the \emph{Stationary Current Regime} (SCR).
In the remaining part of this section, we will show that in this kinematic range it is possible to rigorously approximate 
all ingredients of TPT from statistical averages based on short, non-ergodic trajectories.

We begin by introducing two Green's functions,  $P^{(R)}(x_f, \tau| x, t)$ and $P^{(P)}(x, t| x_i, 0)$, which satisfy the same  FP equation:
\be\label{P*}
\left( \frac{\partial}{\partial t}- \hat H_{FP}\right) P^{(k)}(x,t|x_i,0) = \delta(x-x_i) \delta(t)\qquad (k= R, P), 
\ee
but they obey different boundary conditions. In particular,  $P^{(R)}(x_f, \tau| x, t)$ vanishes at the boundary of the reactant state, i.e.
$  \left. P^{(R)}(x,t|x_i,0)\right|_{x\in \partial R}=0$, while $P^{(P)}(x_f, \tau| x, t)$ vanishes at the boundary of the product state, 
$  \left. P^{(P)}(x,t|x_i,0)\right|_{x\in \partial P}=0$. 

These Green's function can be used to construct the following distributions:
\be\label{Q_R}
Q^{(R)}(x,t) &=& \int dx_f  ~h_P(x_f)~ P^{(R)}(x_f, t_f| x, t_f-t)\\
\label{Q_P}
Q^{(P)}(x,t) &=& \int dx_f  ~h_R(x_f)~ P^{(P)}(x_f, t_f| x, t_f-t),
\ee
where we have introduced some intermediate time   $t\in [0, t_f]$. 

Let us now perform the spectral representation of  the Green's function $P^{(R)}(x_f,t_f|x,t_f-t)$ which enters Eq. (\ref{Q_R}) using the left eigenmodes of the FP operator:
\be\label{spectralQ}
Q^{(R)}(x,t) &\equiv&  \sum_n ~r^{(R)}_n~ L^{(R)}_n(x)~ e^{-k_n t} \qquad \textrm{where}\\
\hat H_{FP}^\dagger ~L_n^{(R)}(x) &=& k_n~L_n^{(R)}(x)\\
r_n^{(R)} &=& \int dx_f ~h_P(x_f)  R_n^{(R)}(x_f).
\ee

We note that, due to the presence of an absorbing boundary condition at $\partial R$, the lowest eigenvalue $k_0$ in this spectral  decomposition is of order $k_S$.  Indeed, the probability of observing the particle slowly decays with time at a  rate $\sim k_S$, because trajectories are ``annihilated" any time the system reaches the reactant state. All other eigen-frequencies in the spectrum are order $k_F$.  

We note also that the lowest right eigen-mode $R^{(R)}_{0}(x)$ is locally Gibbsean and exponentially suppressed outside  the product state $P$, i.e. 
\be
R^{(R)}_{0}(x) \simeq \frac{e^{-\beta U(x) }}{Z_P}~ h_P(x), \qquad \left(Z_P = \int dx ~e^{-\beta U(x) }~ h_P(x) \right) 
\ee
thus $r^{(R)}_0\simeq 1$.  Conversely, the slowest left-eigenmode $L^{(R)}_{0}(x)$ is nearly uniform within the product state $P$ and gradually decays in the transition region until it vanishes at the boundary of the reactant.  The  orthonormality condition:
\be
\int d x R^{(R)}_{0}(x)~ L^{(R)}_{0}(x) = 1 
\ee
implies that $L^{(R)}_{0}(x)$ must  approach $1$ at the border of the $P$ state.  

If  the time interval $t$ entering Eq. (\ref{spectralQ}) is chosen in the SCR, i.e. it obeys the inequalities (\ref{SCR}), 
then
\be\label{QRapp}
Q^{(R)}(x,t) \simeq L_0^{(R)}(x).
\ee
The same derivation can be repeated for $Q^{(P)}(x,t)$ leading to  $Q^{(P)}(x,t) \simeq L_0^{(P)}(x)$.

Finally, let us introduce the time integrals of the  $Q^{(P)}(x,t)$ and  $Q^{(R)}(x,t)$ distributions:
\be\label{q+SCR}
q_{SCR}^{+}(x) &=& \frac{1}{t_f-\tau_0}~ \int_{\tau_0}^{t_f} dt ~Q^{(R)}(x,t) \\
\label{q-SCR}
q_{SCR}^{-}(x) &=& \frac{1}{t_f-\tau_0}~ \int_{\tau_0}^{t_f} dt~ Q^{(P)}(x,t) .
\ee
Here $t_f$ and $\tau_0$ are chosen in such a way to restrict the dynamics in the SCR. In particular:
\be
(\tau_0 k_F)^{-1}\ll1\qquad t_f k_S \ll 1. 
\ee
With this choice, 
\be\label{q^+_SCR}
q_{SCR}^{+(-)}(x) \simeq L_0^{R(P)}(x),
\ee
thus
\be\label{bckfp}
D \left( \nabla^2 -  \beta  \nabla U(x) \nabla \right) q_{SCR}^{+(-)}(x) \simeq 0 .
\ee
Again, corrections to the right-hand-side are  $\mathcal{O}(k_S t_f)$. 
We also note that  $q_{SCR}^{+(-)}(x)$ obeys the same boundary conditions of the forward (backward) committor. Thus, it provides an estimate of the $q^{+(-)}(x)$ functions defined in  TPT (see appendix \ref{TPTsect}). 


Let us now discuss the analog estimate of  the transition probability density $m_{T}(x)$.
The density of points visited by the reactive paths which perform the barrier-crossing transition in the SCR is given by the distribution
\be
\label{m_SCRdef}
m_{SCR}(x) \equiv \frac{1}{t_f-\tau_0} \int_{\tau_0}^{t_f} dt \int d x_i \int d x_f~ h_P(x_f) P^{(R)}(x_f, t_f| x, t) P^{(P)}(x,t| x_i, 0) \rho_0(x_i)~h_R(x_i),
\ee
where $\rho_0(x)$ is the local Gibbsean distribution introduced in Eq. (\ref{localGibbs}). 
We emphasize that the boundary conditions on the Green's functions ensure that the  paths contributing to the $m_{SCR}(x)$ distributions  are  reactive, i.e. do not backtrack to the $R$ state, up to exponentially suppressed contributions. 

To show that the definition in Eq. (\ref{m_SCRdef}) provides the correct approximation to the TPT transition path density $m_T(x)$, we shall prove that it can be expressed as a product between the Gibbs distribution and the SCR estimate forward- and backward- committors, as in Eq.  (\ref{TPTprop1}). 
Introducing the distribution
\be
 \label{P_P}
P^{(P)}(x,t)  &\equiv& \int dx_i ~ P^{(P)}(x, t| x_i, 0)~\rho_0(x_i), 
\ee
the density in Eq. ({\ref{m_SCRdef}) reads

\be\label{mT}
m_{SCR}(x) &=&  \frac{1}{t_f-\tau_0} \int_{\tau_0}^{t_f} ~dt ~Q^{(R)}(x,t_f-t) P^{(P)}(x,t).
\ee
Using the detailed balance condition, the   we find
$
P^{(P)}(x,t) =~e^{-\beta U(x)} \frac{1}{Z_R}~Q^{(P)}(x,t).
$
Then, inserting this result into Eq. (\ref{m_SCRdef})  we find
\be\label{m_SCR}
m_{SCR}(x)&=& \frac{e^{-\beta U(x)}}{Z_R~{(t_f-\tau_0)}} \int_{\tau_0}^{t_f} ~dt Q^{(R)}(x,t_f-t) Q^{(P)}(x,t).\nn
\ee

Finally, recalling that $Q^{(R)}(x,t)$ and $Q^{(P)}(x,t)$ are nearly time-independent  in the SCR, and using Eq.s (\ref{q+SCR}) and (\ref{q-SCR}) we recover a fundamental result of  TPT (cfr. Eq.  (\ref{TPTprop1}) in appendix A): 
\be\label{mscr}
m_{SCR}(x)&\propto & e^{-\beta U(x)}~ q^+_{SCR}(x)~(1-q^+_{SCR}(x)~). 
\ee

Within the same framework it is possible to define the reactive current in the SCR in complete analogy with Eq. (\ref{m_SCRdef}):
\be\label{J_SCR}
J^i_{SCR}(x) &=&\frac{-D}{t_f-\tau_0}\int_{\tau_0}^{t_f} ~dt Q^{(R)}(x,t_f-t) (\stackrel{\rightarrow}{\nabla}-\stackrel{\leftarrow}{\nabla} + \beta \nabla U(x) ) ~P^{(P)}(x,t).
\ee
In this equation the symbols $\stackrel{\rightarrow}{\nabla}$ and $ \stackrel{\leftarrow}{\nabla}$ denote the gradient operator acting on the right and on the left, respectively. Note that the term $-\stackrel{\leftarrow}{\nabla} $ is required to account for  recrossing contributions. 
Again, the detailed balance condition implies
\be\label{J_SCR2}
J^i_{SCR}(x) = D \, \nabla^i q^+_{SCR}(x) \frac{e^{-\beta U(x)} }{Z_R} 
\ee
thus recovering a second fundamental result of TPT (cfr.  Eq.  (\ref{TPTprop2}) in appendix A). 

\section{Computing the Committor Function }\label{SCPSsect}

The problem of estimating the forward  committor function $q^+(x)$ using molecular simulations has been widely discussed in the literature and a number of methods have been proposed to exploit the computational advantages of enhanced sampling techniques (see e.g. \cite{Elberq, MSMq, TPSq}). 
In this section, we develop a scheme based on the recently proposed SCPS method \cite{SCPS} (briefly reviewed in appendix \ref{SCPSalg}).  The advantage of this  algorithm is that it promises to be applicable to large systems. For example, it has been validated against plain MD folding simulations performed on the Anton supercomputer~\cite{anton} and it was applied to simulate the folding of a 130 residues protein, which folds in the seconds timescale. These results were obtained in just a few days using a small computer cluster (consisting of about 100 cores). 

The SCPS algorithm  can be formally derived  from the path integral representation of the Langevin equation, by performing a mean-field approximation over some auxiliary variable. 
Such an approximation turns the original Langevin dynamics into a special type of rMD: Indeed, it gives raise to a history dependent biasing force, which acts along the direction tangent to the mean-transition pathway and switches on only when the system tries to backtrack towards the reactant. On the other hand, the biasing force remains latent when the system spontaneously progresses towards the product. The mean transition paths obtained in this way  provide a self-consistent approximation to the principal curves of the original Langevin dynamics. 

We stress that in the original applications of rMD schemes to protein folding \cite{rMD2, PNAS1, BFA}  and to conformational transitions \cite{PNAS2}, the  direction of the biasing force needed to be  specified \emph{a priori}.  In contrast, in the SCPS approach the reaction coordinate is computed self-consistently, through an iterative procedure. While a few  enhanced path sampling approaches based  on a self-consistently evaluated reaction coordinate  have already been proposed in the literature (see e.g.  \cite{metaSC,PATHmeta}),  the high computational efficiency of the rMD scheme makes  the SCPS approach applicable to   large and complex reactions, such as protein folding.  

A main  drawback of the SCPS approach (and in general all  methods based on the rMD) is that its microscopic dynamics is not reversible. Consequently, extracting kinetic and thermodynamic information from rMD trajectories is in general  non-trivial. 

In the following,  we propose a method based on SCPS to obtain a full foliation of the configuration space in terms of iso-committor hyper-surfaces. We stress that the surfaces obtained in this way are not restricted to the hyperplanes, but can have arbitrary shape. The average paths calculated using the SCPS algorithm starting from $N_R$ different initial conditions can be combined in order to define a global collective variable  introduced in Ref.  \cite{tubevar}: 
\be\label{sigma}
\hspace{-0.6 cm} \sigma(x) \equiv \frac{1}{N_R~t_f}\sum_{k=1}^{N_R} \frac{\int_0^{t_f} dt' t' e^{-\lambda~|| x(x) -\langle  x(t')\rangle_{(k)} ||^2}}{\int_0^{t_f} dt' e^{-\lambda ||x - \langle x(t') \rangle_{(k)} ||^2}}.
\ee   
In this equation, $||\ldots||^2$ denotes some suitable norm, while $\langle x(t')\rangle_{k}$ is the instantaneous mean-value of $x$ calculated  over all the transition  pathways generated from the $k$-th initial condition.  
We note that in the large $\lambda$ limit the function $\sigma(x)$ assigns to the configuration $x$ the time label of the closest frame among all the frames in the $N_R$ mean-paths (see Fig. \ref{sigmaFig}). This number is normalized by the total number of frames in the reactive trajectories, thus $\sigma \in [0,1]$.
In particular, $\sigma \sim 0$  near the reactant, while  $\sigma\sim 1$ near the product.  


The question arises whether $\sigma(x)$ can be regarded as a good reaction coordinate, i.e. provides the same foliation of configuration space of the committor function. In appendix \ref{q_param_sigma}, we explicitly prove that in the vicinity of the principal curve iso-$q^+$ and iso-$\sigma$ hyperplanes coincide. In the following, we shall assume that this property holds throughout the whole kinetically relevant region of configuration space and define a computationally efficient procedure to compute $q^+(x)$ from $\sigma(x)$.

\begin{figure}[t!] 
\center 
\includegraphics[width=0.5\textwidth]{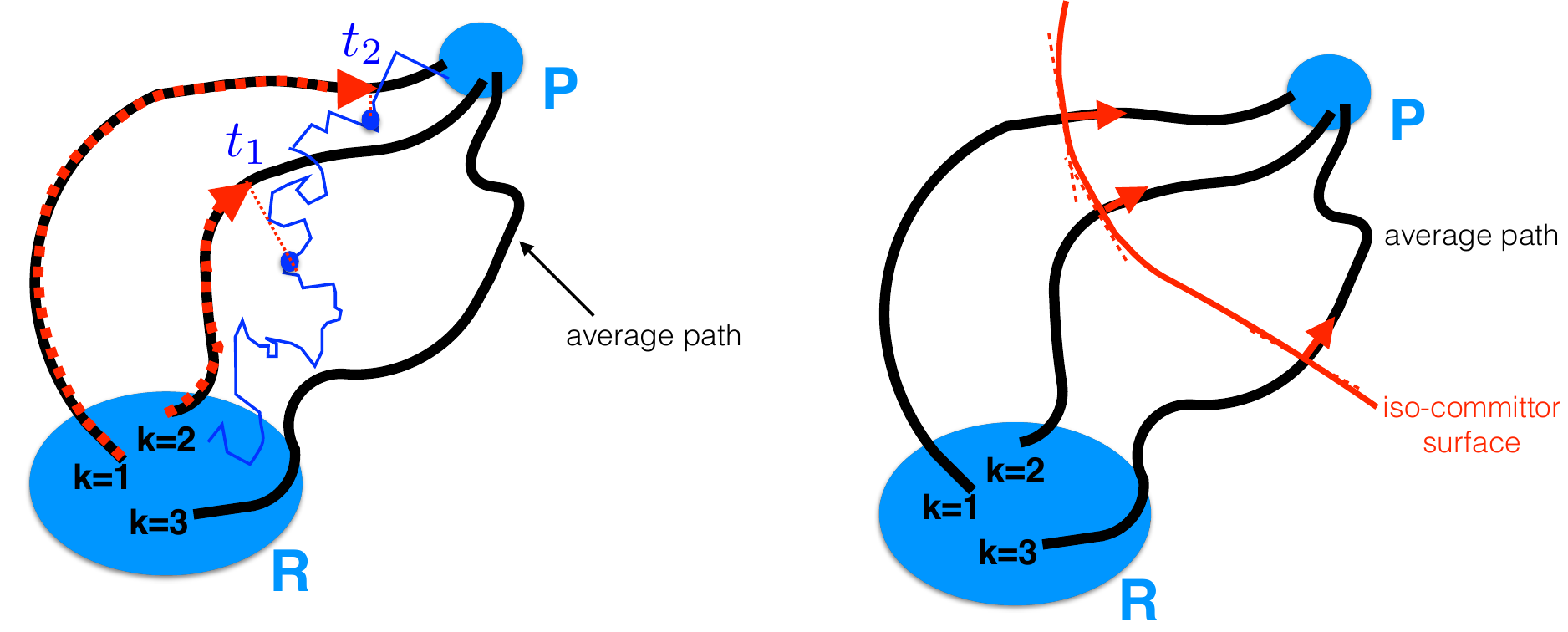} 
\vspace{0mm}
\caption{Left panel: Geometric interpretation of tube variable $s_\lambda(x)$ (and its multi-path generalization  $\sigma(x)$). Right panel:  Schematic representation of vectors tangent to the average transition pathways, which are locally orthogonal to the iso-committor hypersurface. }\label{sigmaFig} 
\end{figure}

To construct such a map, we recall  that the  committor functions obey the backward Kolmogorov equation, Eq. (\ref{bckfp}).   Since $\nabla \sigma$ is parallel to $\nabla q^+$ the partial derivatives of $q^+(x)$ with respect to all directions perpendicular to $\nabla \sigma$ are null, thus it is sufficient to enforce the Kolmogorov equation along the one-dimensional manifold provided by the mean path. In conclusion, the values of  $q^+(x)$ computed along the frames of any of the $N_R$ average reaction trajectories  obey the following equation: 
\be
\frac{1}{\Delta d_{l+1,l}}\left( \frac{q^+_{l+1}-q^+_l}{\Delta d_{l+1, l}} -  \frac{ q^+_{l} - ~q^+_{l-1}}{\Delta d_{l,l-1}} \right) - \frac{\beta}{\Delta d_{l+1,l}} \left(q^+_{l+1} - q^+_{l}\right) \hat{n}_l\cdot \langle \nabla U_l\rangle \simeq 0. 
\ee
In this equation, the index $l$ labels the different frames of the average transition path, i.e. $q^+_l\equiv q^+(\langle x(t_l)\rangle$, $\hat n_l$ is the unit vector in configuration space  tangent to the principal curve at the $l-$th frame and $-\langle \nabla U_l\rangle $ is evaluated by averaging over the force in the $l$-th frame of the configurations used to compute the average transition path. Finally, $\Delta d_{l+1,l}$ is the  Euclidean distance between the $l$-th and $l+1$-th frames of the average path, $d_{l+1, l}\equiv \sqrt{|| \langle x(t_{l+1})\rangle - \langle x(t_{l+1})\rangle||^2}$.

Based on this observation, our best estimate of the assignment of values of $q^+$ to the iso-$\sigma$ hyper-surfaces is obtained by minimizing the following function of the  $(q^+_1, \ldots, q^+_{N_t})$ variables:
\begin{equation}\label{minimization}
I =\sum_{l=1}^{N_t}\left[\frac{1}{\Delta d_{l+1,l}}\left( \frac{q^+_{l+1}-q^+_l}{\Delta d_{l+1, l}} -  \frac{ q^+_{l} - ~q^+_{l-1}}{\Delta d_{l,l-1}} \right) - \frac{\beta}{\Delta d_{l+1,l}} \left(q^+_{l+1} - q^+_{l}\right) \hat{n}_l\cdot \langle \nabla U_l\rangle \right]^2.
\end{equation}
The numerical optimization of this function returns the value of $q^+$ attained at any given iso-$\sigma$ hyper-surface, i.e the function $q^+(\sigma)$. 

\section{Sampling of  the Transition Region}\label{CLDsect}
In this section, we discuss two different  computational schemes to efficiently sample the transition region. Both these algorithms capitalize on the calculation of $q^+(x)$ in order to avoid  simulating thermal oscillations in the reactant and product states. 
On the other hand, they provide slightly different information and have relative advantages and disadvantages. They represent a generalization to the continuum case of sampling schemes originally introduced by Vanden Eijnded and co-workers in the framework of discrete Markov processes \cite{CLDMSM1, CLDMSM2}.

The first of these algorithms is based on  modifying the original Langevin equation in order to include a position-dependent biasing force. The corresponding 
 dynamics can be used to define a non-equilibrium  stochastic process with a stationary solution given by the $m_T(x)$ distribution. 

In high-dimensional  rugged energy landscapes, however,  the sampling of the transition density distribution by this algorithm may still be computationally demanding. This is because reactive paths can significantly detour from the main productive reactive channels, by performing loops or getting stuck in kinetic traps (a discussion of this point is given in Ref.s \cite{CLDMSM1, CLDMSM2}). 

To overcome this problem we consider a second algorithm, which enables one to sample the Boltzmann distribution restricted to the transition region by generating trajectories which travel only forward along the committor, thus avoiding detours.  The main shortcoming of this second approach is that it involves a \emph{history dependent} biasing force which breaks microscopic reversibility. Consequently, the trajectories generated this way do not have a direct physical interpretation in terms of reactive events.

\subsection{Sampling $m_T(x)$ by Conditional Langevin Dynamics}\label{CLDalg}

The transition path density distribution $m_T(x)$ can sampled by integrating the following modified Langevin equation:
 \be\label{LB2}
\dot x &=&  D  \left(-  \beta \nabla U (x) +  2 \beta^{-1} \frac{\nabla q^+(x)}{q^+(x)} \right) + \eta(t). 
\ee
This equation was  first derived in Ref. \cite{CLD0} within the general framework of  diffusion processes, by means of the Doob transform \cite{Doob}.  Vanden-Eijnden and co-workers also obtained an analog result within the framework of the theory of Markov jump processes \cite{CLDMSM1, CLDMSM2}.

In appendix D, we independently recover Eq. (\ref{LB2}) starting directly from the definition of transition path density. Using the  Conditional Langevin Dynamics (CLD) formalism ---for recent applications see also Ref.s \cite{CLD1, CLD2}--- and exploting the properties of the SCR regime, we show that the probability $\mathcal{P}(x,t)$  for a transition pathway  to visit the point $x$ at some time $t\in [0,t_f]$ obeys the modified Forkker-Planck equation: 
\be\label{PDE}
\frac{\partial}{\partial t} \mathcal{P}(x,t) &=&  D\nabla\cdot \left( \nabla + \beta \nabla U(x) - 2 \nabla \ln q^+(x)  \right) \mathcal{P}(x,t) 
\ee
It is immediate to verify that $m_T(x) \propto q^+(x) (1-q^+(x)) e^{-\beta U(x)}$ is a stationary solution of this equation, with a non-vanishing probability current $J_T(x) \propto \nabla q^+(x) e^{-\beta U(x)}$. 
 
The Langevin equation associated to this modified Fokker-Planck Equation is precisely Eq. (\ref{LB2}). However,  it is important to note  that integrating Eq. (\ref{LB2})  to generate ergodic trajectories would not lead to the correct transition path density. Instead, it would lead to an equilibrium distribution   $\propto (q^+(x))^2~\exp[-\beta U(x)] $, which has no physical relevance. In order to compute $m_T(x)$ from Eq. (\ref{LB2})  one must consider a non-equilibrium process in which many short independent trajectories are  generated in the reactant  and terminated as soon as they reach the product.

Finally, we emphasize that Eq. (\ref{LB2}) is useless, unless the committor function has been previously calculated. Using  SCPS to estimate $q^+(x)$, we obtain the following expression for the biasing force entering Eq. (\ref{LB2}):
\be\label{BF}
 F^i_{CLD}(x) = 2 \beta^{-1} \frac{\nabla^i q^+(x)}{q^+(x)} = 2\beta^{-1} \frac{dq^+}{d\sigma}~(q^+[\sigma(x)])^{-1}~\nabla^i \sigma(x),
 \ee 
 where the field $\sigma(x)$ and the function $\frac{dq^+}{d\sigma}$ are obtained from SCPS using the procedure described in section \ref{SCPSsect}. 
This bias for   is ideal in the sense that it yields the same transition path ensemble of the original Langevin, in the limit in which the committor function  is exactly known and in the SCR.

\subsection{Sampling the Boltzmann Distribution in the Transition Region by Ideal rMD}\label{irMDalg}
Let us consider a special kind of rMD simulation,  in which   the committor is used  as biasing collective coordinate. Namely, we introduce the following history-dependent  force into the equations of motion,
\be\label{irMD}
F^i_{irMD}(x, t) =  k_R \nabla q^+(x)~\theta(q^+_M(t) - q^+(x))~\xi\left( q^+_M(t) - q^+(x)\right).
\ee 
In this equation,  $q^+_M (t)$ is the maximum value attained by the committor up to time $t$ and the function $\xi(x)$ is non-negative for $x>0$. We note that the specific functional form of $\xi(x)$ only affects the dynamics when the system attempts to backtrack toward the reactant. Thus, in the strong bias limit $k_R\to \infty$  in which backtracking parts of the reactive paths are strongly quenched, the specific choice of $\xi(x)$ becomes irrelevant.
We refer to the biased dynamics defined by Eq. (\ref{irMD}) as to the  \emph{Ideal rMD} (irMD). 
In analogy with Eq. (\ref{BF}),  the mapping between the $\sigma(x)$ and $q^+(x)$ fields discussed in the previous section  enables Eq. ~(\ref{irMD}) to be expressed  in terms of quantities which are calculable by SCPS.

It is important to emphasize, however,  that the irMD is not equivalent to a plain SCPS simulation. First of all, the SCPS algorithm involves \emph{two} biasing forces defined along different coordinates, $s_\lambda$ and $w_\lambda$ (see appendix \ref{SCPSalg}). The bias along $w_\lambda$  is introduced to ensure that  the paths  travel close to the average path calculated at the previous SCPS iteration. This  is done to increase the computational stability of the algorithm, ensuring that the update of the principal curve during subsequent iterations is done in a  quasi-adiabatic way.  

We also stress that the mapping between the $q^+(x)$ and $\sigma(x)$ fields is in general non-linear. In particular,  $\sigma(x)$  typically increases in a nearly monotonic way along the straight line connecting the bottom of different metastable states, while $q^+(x)$ remains nearly constant throughout the metastable basins. This is expected, since different points in the same  basin are kinetically close, thus  have similar values of the committor. As a result, in a irMD simulation the strength of the biasing force is small inside  such regions, reducing the amount of steering and enabling a more exhaustive exploration of these states. 

In appendix \ref{proveirMD} we show that the irMD can be used sample the Boltzmann distribution restricted to the reactive region, by generating nonequilibrium Langevin trajectories which proceed only forward along the direction set by the committor. This result generalizes the no-detour path dynamics introduced in Ref.s~\cite{CLDMSM1, CLDMSM2} in the framework of discrete Markov jump processes. 

\section{An Illustrative Application}\label{example}

In this section, we illustrate our  scheme to compute the committor function and sample the transition region in a simple two-dimensional problem. In particular, we consider the  diffusion of a particle on the two-dimensional energy surface (shown as background in  the upper left panel of Fig. \ref{fig:m_SCR}):
\begin{equation}\label{potential}
\begin{split}
U(x,y) &= u_0\left\{e^{ -(x^2 + y^2)} - \frac{3}{5} e^{ - (x^2 + \left( y - b_0 \right)^2)} - e^{ - (( x - x_0)^2 + y^2)} -e^{ -( (x+x_0)^2 + y^2)} \right\} + w_0\left[ x^4 + \left( y - a_0 \right)^4\right]
\end{split}
\end{equation}
 where $x_0=1$, $a_0=\frac{1}{3}$, $b_0= \frac{5}{3}$,  $u_0=5$ and $w_0= \frac{1}{5}$ in the appropriate units. This is a simplified version of the potential studied in Ref. \cite{TPT2}, which contains three minima 
centered at  \( \textbf{x}_R =(x_R, y_R) =(-1,0) \),  \(\textbf{x}_I =(x_I, y_I)= (0, 1.5)\) and \( \textbf{x}_P=(x_P, y_P) = (1,0)\), respectively. We defined the reactant $R$ and the product $P$ basins as the regions around \(\textbf{x}_R\) and \(\textbf{x}_P\), respectively,  where \(U(x,y) \leq -2.5 k_B T\). At low temperature,  the transition from $R$ to $P$ occurs through trajectories which visit the intermediate metastable state,  $I$, centered around \(\textbf{x}_I\). In all simulations, the thermal energy was chosen to be $k_BT= 0.15$.

The $R$ to $P$ reaction can be investigated by integrating the standard Langevin equation (\ref{Lang}), and using an integration time step $dt =0.02$ (in units of inverse diffusion coefficient $\gamma$).  Due to the high energy barriers separating the states, even for such a simple system the complete characterization of the reaction by plain Langevin simulations is rather computationally expensive. However, by combining  the algorithms defined in section \ref{SCPSsect} and \ref{CLDsect}, this computational cost can be drastically reduced. First, we discuss the calculation of the committor function $q^+(x)$ based on the SCPS algorithm. Then, we show how the CLD and the irMD algorithms introduced in the previous section capitalize on this information to yield an accurate sampling of the $m_T(x)$ distribution. 

\subsection{Generation of Transition Pathways by SCPS}
The first step towards computing the committor using the SCPS method consists in generating an ensemble of reactive pathways. To this goal, we implemented the algorithm described in appendix \ref{SCPSalg}. First, we  performed \(1000\) plain rMD simulations, starting from $\textbf{x}_R$ and biased along the  coordinate 
\be\label{ztoy}
z(\textbf{x}) = \sqrt{(x-x_P)^2+(y-y_P)^2},
\ee
 which measures the instantaneous Euclidean distance  to the product state. The ratchet elastic constant  in Eq. \eqref{rmd_force} was set to $k_R=50$.

With this choice of  collective coordinate and  parameters, all the rMD trajectories reached the product basin within the total simulation time of $4\times 10^3$ time steps. However, the results of this   rMD simulation is flawed by  systematic errors due to the suboptimal choice of the biasing  coordinate.  Indeed, the collective coordinate $z$  ignores the existence of the intermediate state. Moreover, we note that the modulus of the bias force is  very large, approximately twice  that of  the physical force. Both such choices were made  because we were interested to study to what extent the SCPS iterations can correct for systematic errors on the initial trial guess.   

The results  reported in Fig. \ref{fig:m_SCR} show that  the rMD trajectories reproduce at the qualitative level some of the main features of the transition path ensemble, in spite of the fact they were performed using a large  biasing force, acting  along a rather bad  reaction coordinate. In contrast, a plain steered MD with  external force $F_{B}= -k_2 \nabla z(x)$ of comparable magnitude would yield completely wrong information about the reaction mechanism.
However, several systematic errors can be  noticed in the rMD results: first, the heat map showing the density of points is clearly not symmetric, thus it does not reflect the structure of the underlying energy landscape; Moreover,  the presence of an intermediate energy minimum is not evident, as the trajectories do not significantly populate the region around \( \textbf{x}_I\);  Finally,  the average pathway does not cross the intermediate state $\textbf{x}_I$. 

Next, we used the rMD results as the starting point to perform three iterations of the SCPS algorithm.  At each iteration, we first computed the average path $\langle {\bf x}(t) \rangle$ using the reactive trajectories generated at the previous iteration. Then, we used this path to define two collective coordinates in Eq.s \eqref{s_lambda} and \eqref{w_lambda} with $t_f = 4\times 10^3 dt$ and $ \lambda = 30$. 
Details about the selection of the reactive part of the trajectories,  the averaging procedure and the choice of the $\lambda$ parameter are provided in the Supplementary Material. 
At each iteration we ran 5000 independent rMD simulations employing  the bias force defined in Eq. \eqref{scps_force}. After 3 iterations we observed that the average path does not appreciably change, according to the $\mathcal{L}_2$ norm ( the results are reported  in the Supplementary Figure 1)

\begin{figure}[t!]
\centering
\includegraphics[width=0.6\textwidth]{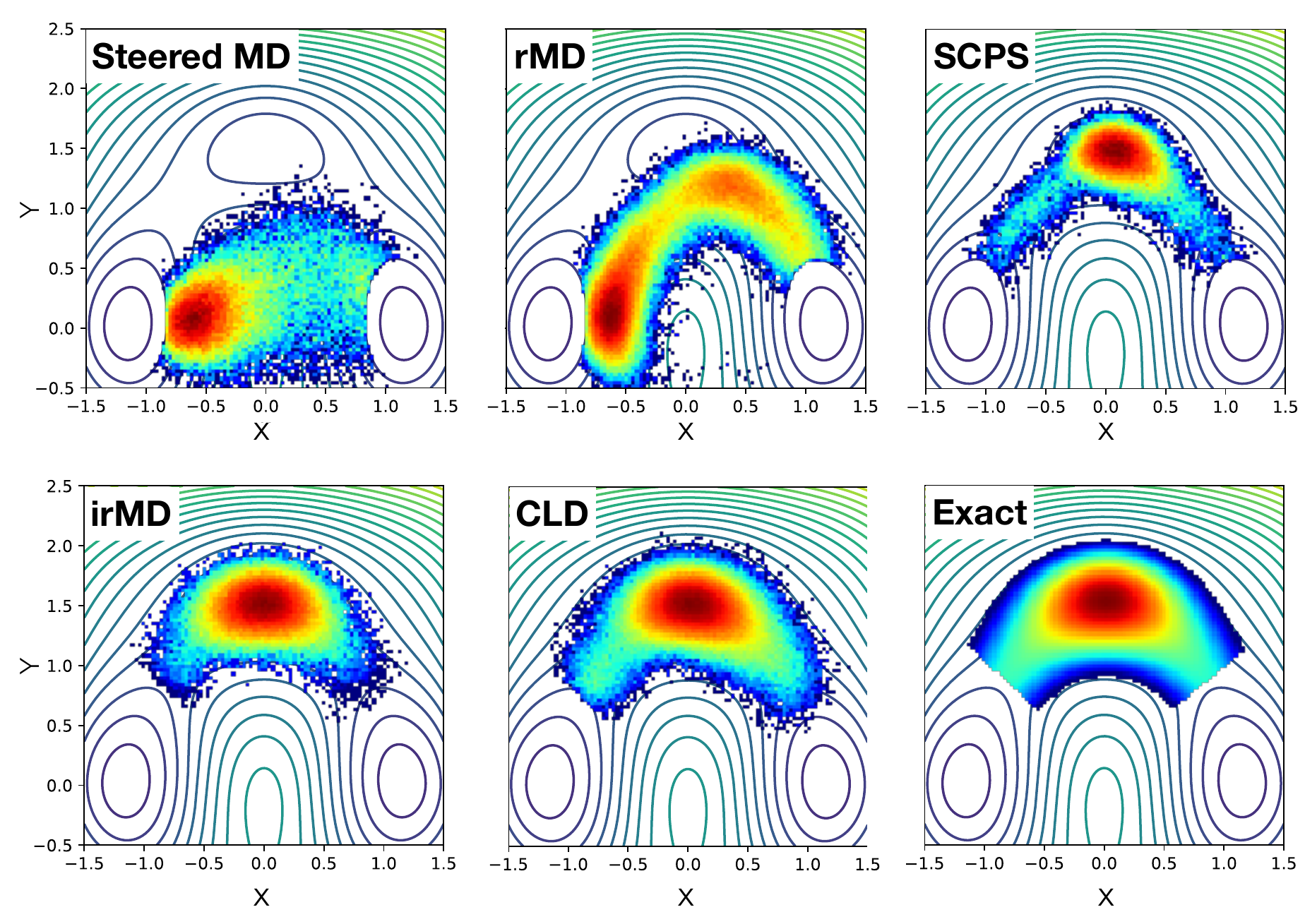}
\caption {Transition  probability density for our two-dimensional toy model, computed with different methods. The background in the upper left panel shows the countour plot of the energy surface.}
{\label{fig:m_SCR}}
\end{figure}

\subsection{ Computing the Committor Function}

We used the average path obtained after  three SCPS iterations to compute the collective coordinate $\sigma(x)$  in the region  $x \in [-1.5,1.5]$ and $y \in [-0.5,2.5]$. In Fig. \ref{fig:committor} we compare (a) this result with (c) the committor $q^+(x)$ computed by plain Langevin simulations using the scheme reported in the Supplementary Material. We can see that these two plots agree at the qualitative level: indeed both iso-committor and iso-$\sigma$ lines are approximatively parallel in the transition regions between the $R$ and $I$ and between $I $ and $P$. Furthermore, both the iso-$q^+$ and the iso-$\sigma$ surfaces bend near the region around $x\sim (0,0)$, signalling the existence of a high barrier, which is never overcome at this temperature. Notice that, in contrast, this  region is often visited by the trajectories generated by steered MD.  Overall, these results imply that iso-$\sigma$ lines generated by SCPS are reasonable  approximations of iso-committor lines. 

On the quantitative level, there are some significant differences between the exact calculation of $q^+(x)$ obtained by plain Langevin simulations and the numerical values of $\sigma(x)$ obtained via SCPS. In particular, it is clear that the  spacing between iso-$\sigma$ lines is more pronounced than that between the exact iso-committor lines. 
This  signals the existence of a non-linear mapping between $q^+(x)$ and $\sigma(x)$. 

To obtain an estimate of $q^+(x)$ from the iso-$\sigma$  surfaces evaluated with SCPS we applied the  technique introduced in section \ref{SCPSsect}. The details of the minimization procedure are discussed in the Supplementary Material. The starting value of the functional was $I\sim 1.5\times 10^3$, while after the minimization the value of the functional dropped to $I\sim 10^{-3}$. This means that the resulting estimation of the committor satisfies the Kolmogorov equation with a precision of $\mathcal{O}(10^{-3})$. The results are shown in Fig. \eqref{fig:committor}(c): We note that the spacing between the calculated iso-$q^+$ lines resembles much better the exact result. The most significant difference between $\sigma(x)$ and $q^+(x)$ is shown in Fig. \eqref{fig:committor} (d): indeed $q^+(x)$ exhibits a plateau in correspondence of the intermediate state. A moderate disagreement between the committor obtained by our method and the one obtained from  exact Langevin simulations can be noted only in the regions which are rarely sampled by the  transition pathways.

\subsection{Transition Probability Density and Reaction Tubes}

The SCPS algorithm used to estimate the committor simultaneously produces a mean-field estimate of  the transition path ensemble. 
Using the schemes discussed in section \ref{CLDsect} it is possible to improve on this estimate, and to  sample $m_T(x)$  with an accuracy which is affected only by the error on the SCPS estimate of $q^+(x)$. 

In particular, we recall that the CLD defined in section \ref{CLDalg}  directly provides the $m_T(x)$ distribution, while the  irMD discussed in section \ref{irMDalg}  yields the Boltzmann distribution restricted to the reactive region. From the latter distribution, however,   $m_T(x)$  can be readily obtained by re-weighting according to the reactivity probability factor:  $e^{-\beta U(x,y)} \to q^+(x,y) (1- q^+(x,y)) e^{-\beta U(x,y)}$. 

We performed $10^3$ independent  CLD and irMD simulations, each lasting $4 \times 10^3$ time steps, employing a ratchet constant $k_R=50$. In irMD simulations we chose  $\xi(x)= x$ (cfr. Eq. (\ref{irMD})). Since $q^+(x)$ vanishes at the boundary of the reactant, the initial configurations of these simulations have to be chosen in the reactive region. In our simulations, we chose to initiate both the CLD and the irMD trajectories from the points visited by the trajectories of the last SCPS iteration which lied on the iso-$q^+$ surface around $q^+ = 0.01$.

Results are summarized in Fig. \ref{fig:m_SCR}.   The accuracy of the  different simulation methods can be assessed by comparing with the reference distribution shown in the lowest right panel, which represents the unbiased result. This  distribution was obtained using the algorithm detailed in the Supplementary Material to evaluate the committor.

The basic rMD algorithm fails in detecting the existence of the intermediate state and  introduces a systematic shift in the position of the reactant state. In contrast,  SCPS is able to  reproduce the reactive distribution almost at a quantitative level. However,  the position of the probability peak in  $m_T(x)$  corresponding to the intermediate state is still sensibly shifted to the right.

The heat maps computed by irMD and CLD  are much more accurate,  in spite of a small error in the SCPS estimate of the committor function (cfr. Fig. \ref{fig:committor}). Arguably, such a good agreement is a consequence of the fact that the  SCPS estimate of $q^+$  is relatively less  accurate only in a region where $\nabla q^+$ is  small, thus where the biasing force is weak. As result,  the overall effect of the systematic error introduced by our estimate of  $q^+$  is negligible.

To complete the description of the reaction mechanism we computed the reaction current applying Eq. \eqref{J_SCR2}. Then, we constructed the reaction tube using algorithm defined at the end of appendix \ref{TPTsect}. The results in Fig. \ref{fig:tubes} show that the SCPS method is  able to locate the correct reaction channel. Only the bending point of the streamlines inside the intermediate state is slightly shifted to the right, arguably due to a small error in our estimate of the committor function. 

\begin{figure}[t!]
\center 
\includegraphics[width=0.65\textwidth]{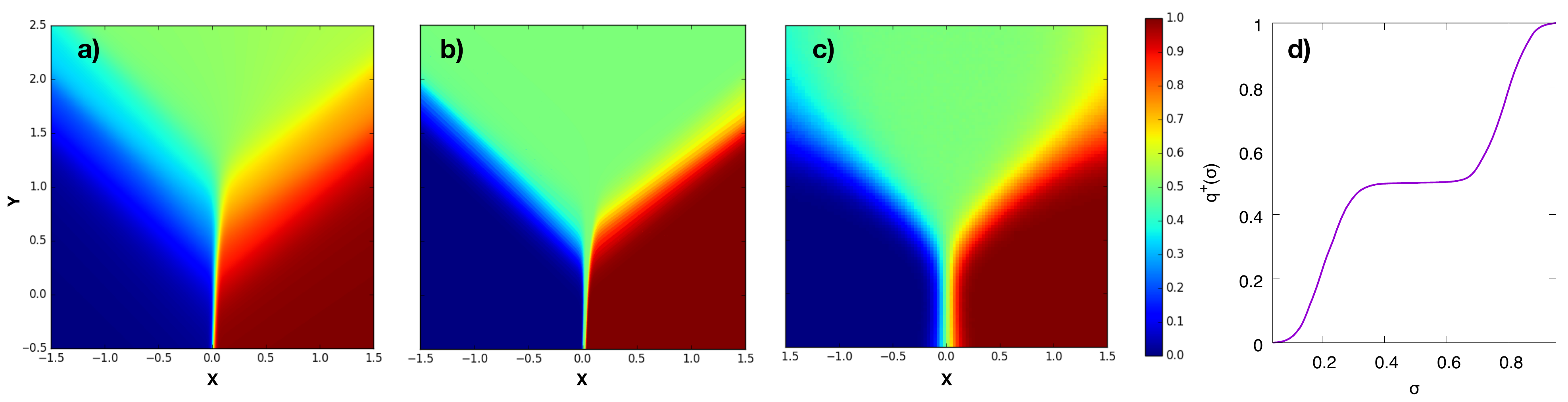}
\caption{(a) $\sigma(x)$ obtained from the third and last iteration of SCPS; (b) $q^+_\text{SCR}(x)$ obtained from the third and last iteration of SCPS after minimizing the functional given in Eq.  (\ref{minimization}); (c) committor estimated in an unbiased way using the procedure described in the Supplementary Material; (d) parametrization $q^+[\sigma(x)]$ obtained using the functional minimization in Eq. \eqref{minimization}. } 
\label{fig:committor}
\end{figure}

\section{Conclusions}\label{conclusions}
In this work we developed efficient computational schemes to   compute all the ingredients of TPT, namely the forward committor function  $q^+(x)$, the transition probability density $m_T(x)$ and the transition current $J_T(x)$. 

TPT is  formulated in terms of statistical averages performed over a long ergodic  trajectory. On the other hand, the computational efficiency of enhanced path sampling techniques  relies on the possibility of generating directly the reactive paths,  i.e.  short nonequilibrium trajectories.
To reconcile theoretical rigor with computational efficiency, in  section \ref{SCRsect}  we have shown that, for  thermally activated transitions,  all the ingredients of TPT can be systematically approximated by averages performed  over such short nonequilibrium trajectories, as long as the time extent of the transition path falls in the time interval (\ref{SCR})  which defines the SCR. 

In sections III and IV, we have tackled the problem of how to efficiently compute the main ingredients of TPT.  First, we have shown how  the committor function $q^+(x)$ can be evaluated by minimizing the functional in Eq. (\ref{minimization}) along the principal curve obtained by means of the  SCPS algorithm. 

Next, we discussed how the knowledge of the committor can be exploited to define two alternative  nonequilibrium schemes to efficiently sample  $m_T(x)$,  with relative advantages and disadvantages.
In particular, generating  reactive  Langevin trajectories by introducing the  biasing force  (\ref{BF})  gives rise to  the same transition path ensemble of the original Langevin equation (up to errors associated with the estimate of $q^+$). On the other hand, since the  potential of this biasing force is only logarithmic, the reactive trajectories generated by this dynamics may contain many loops and kinetic detours which significant increase the computational time. 

The second scheme consists in an ideal version of rMD based on the history-dependent  biasing force (\ref{irMD}), again defined in terms of the committor. This algorithm has the advantage to sample the Boltzmann distribution restricted to the transition region while  keeping to a minimum the amount of computational time invested in  generating the reactive pathways. Indeed, a strong biasing force sets in to hinder backtracking along $q^+$, thus avoids kinetic loops. On the other hand, the trajectories generated by this irMD violate microscopic reversibility, thus cannot be directly physically interpreted as reactive events. 

In this first exploratory work, we have illustrated and validated our methods on a simple two-dimensional toy model.  Consequently, an important question to be addressed in future work is to what extent these  schemes can  be useful to characterize  complex molecular transitions using atomistic force fields. In particular, while the SCPS algorithm has already been applied to generate ensembles of transition pathways for structural reactions as complex as protein folding,  the computational efficiency of the irMD remains to be assessed.  

Finally, a relatively straightforward extension of the present work consists in repeating the derivation without assuming the overdamped limit for the Langevin equation.
\section*{Supplementary Material}
See Supplementary Material for details and figures concerning the numerical implementation of the algorithms.  
\acknowledgments
We greatfully acknowledge fruitful discussions with A. Laio, E. Vanden-Eijnden, H. Orland and G. Ciccotti. 
\begin{figure}[t!]
\centering
\includegraphics[width=0.3\textwidth]{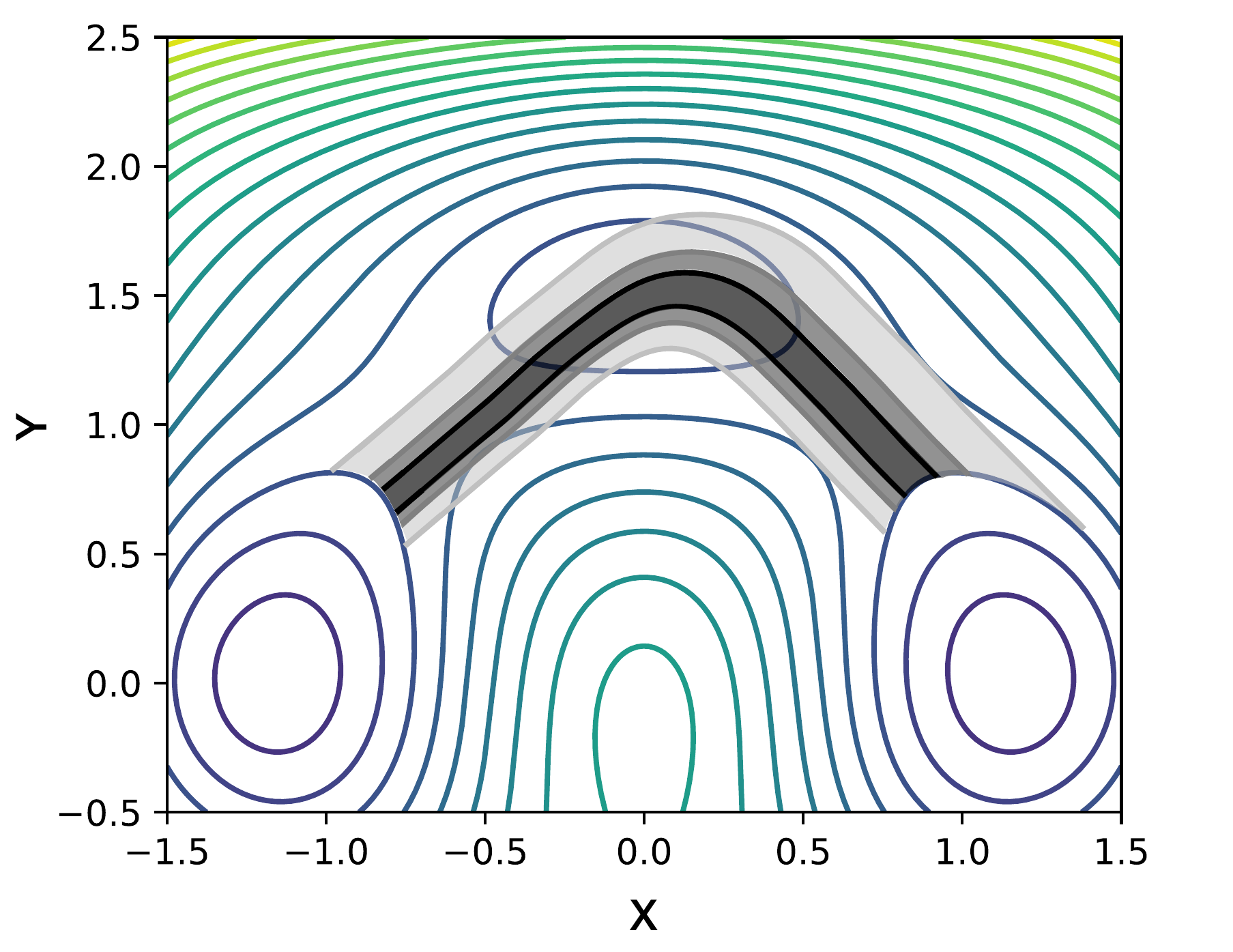}
\caption{Streamlines delimiting the reactive flux tubes corresponding to the \(30\%\) (black), \(60\%\) (dark grey) and \(90\%\) (light grey) of the probability flux.}
{\label{fig:tubes}}
\end{figure}

\appendix
\section{A Brief Review of TPT}\label{TPTsect}

In this section we briefly review TPT  in its original formulation, due to Vanden-Eijnden and co-workers (for  reviews see \cite{TPTreview} and  \cite{TPTnotes}). We consider the general problem of  characterizing the classical  thermally activated transition between some reactant state $R$ and product state $P$,  embedded in a configuration space $\Omega$.
The underlying microscopic dynamics is assumed to obey the following properties: (i) Markovianity (in configuration or phase space) and (ii) microscopical reversibility. The latter request implies ergodicity with respect to some equilibrium distribution. However,  for sake of simplicity, in the present work we specialize on systems obeying the overdamped Langevin dynamics.

The ensemble of  transition paths is defined by harvesting all the reactive segments in  an infinitely long ergodic trajectory $x(t)$ (for the rigorous mathematical definition we refer the reader to \cite{TPTreview}) 
Any generic point $x$ of configuration space  can be  visited by both reactive and non-reactive parts of the ergodic trajectory. 
The probability $P_{\textrm{react}}(x)$ that the ergodic trajectory visits  point $x$ while being reactive can be conveniently expressed by introducing the  forward committor  $q^+(x)$. This  function measures the 
probability that a trajectory initiated at some point  $x\in \Omega_T$  will  enter  the product state $P$  before returning to  the reactant $R$.  Thus,  by definition, $q^+(x)$ obeys  the  boundary conditions $\left. q^+(x)\right|_{ x\in \partial R}= 0$ and $
\left. q^+(x)\right|_{x\in \partial P}= 1$, 
where $\partial R$ and $\partial P$ denote the boundaries of the reactant and product regions, respectively.
Similarly, one can define the backward committor $q^-(x)$ as the probability that a trajectory passing through point $x$ will next return to the reactant $R$ before landing into the  product $P$. By microscopic time-reversibility, $q^-(x)$ can be expressed in terms of the forward committor: $ q^-(x)  = 1 - q^+(x)$.

It can be shown that $q^+(x)$ and $q^-(x)$ obey the so-called stationary backward Kolmogorov equation (see e.g. \cite{TPTnotes}):
\be\label{BKE}
 D \left( \nabla^2 - \beta \nabla U(x) \cdot \nabla \right) f(x) = 0, \qquad f= q^+, q^-. 
\ee
Finally, the  $P_{\textrm{react.}}(x)$  probability is written as $P_{\textrm{react.}}(x) = q^+(x) ~q^-(x)$, thus
$
P_{\textrm{react}}(x) = q^+(x) \left(1- q^+(x) \right).
$

An iso-committor hypersurface is defined as the subset of configuration space  $\partial S_{\bar{q}}$ over which the forward committor is uniform, i.e.   $\left. q^+(x)\right|_{\partial S_{\bar{q}}} = \bar{q}$, with $\bar{q}\in (0,1)$. These manifolds provide a particularly useful foliation of the configuration space.  Indeed, the probability that a reactive trajectory crosses an iso-committor surface at some point $x$  coincides with the equilibrium probability restricted to this surface  \cite{TPT1}:
\be
p_{\partial S_{\bar{q}}}(x) = \frac{1}{\mathcal{Z}_{\partial S_{\bar{q}}}}~e^{-\beta U(x)} \qquad \mathcal{Z}_{\partial S_{\bar{q}}}= \int_{\partial S_{\bar{q}}} d\sigma ~e^{-\beta U(x)}.
\ee
 We emphasize that this result establishes a highly non-trivial relationship between probability densities defined in equilibrium and dynamical conditions, identifying the iso-commitor function as the ideal reaction coordinate. 

The ergodicity assumption of TPT can be exploited to define a scalar time-independent distribution $m_T(x)$ called transition 
path density, which measures the probability  for transition paths to visit a specific configuration $x\in \Omega/(R\cup P)$. 
A central result of TPT consists in relating such a distribution to the committor and equlibrium distribution: 
\be\label{TPTprop1}
m_T(x) = \frac{1}{Z_T} e^{-\beta U(x)}~q^+(x)~(1-q^+(x)),
\ee
where $Z_T$ is a convenient normalization factor: $Z_T = \int_{\Omega_T} dx  \, e^{-\beta U(x)}~q^+(x)~(1-q^+(x))$. 
We emphasize that Eq. (\ref{TPTprop1}) is obtained by exploiting the ergodic assumption.  

As clearly illustrated  in Ref. \cite{TPT2}, $m_T(x)$ alone does not carry information about the reaction kinetics, nor it enables the reaction mechanisms to be univocally identified. Such information  is encoded in the so-called \emph{transition current} $J^i_T(x)$. This vector field is defined in terms of the flux of reactive trajectories going from $R$ to $P$ across an arbitrary hypersurface. 
Like the transition path density $m_T(x)$, also $J^i_T(x)$ can be written in an explicit form which involves the Gibbs distribution and the committor funciton:
\be\label{TPTprop2}
J^i_T(x) = D \nabla^i q^+(x)~\frac{e^{-\beta U(x)}}{Z}.
\ee
Applying the backward Kolmogorov equation ---cfr. Eq.  (\ref{BKE} )--- it is immediate to show that the vector field \( J^i_T \) is divergenceless in \(\Omega_T\). 
Thus, its flux across a dividing hyper-surfaces  is constant. By construction this constant probability flux is simply the \emph{reaction rate}:
\be\label{rate}
k = \sum_{i=1}^{3N}~\int_{\partial{S}} \; J^i_T ~dS_i(x).
\ee
Furthermore, once the current is known it is possible to define a simple algorithm to compute the so-called transition tubes, which carry the information about the reaction mechanism(s): \begin{enumerate}
\item Identify a portion of \(\partial {S}\), which we call \( \partial{A}\), such that the current flux passing through \( \partial{A}\) amounts to certain percentage c.
\item Push forward this surface, evolving each point with the artificial dynamics \be \label{TPTstream}
\frac{d x^i}{d \tau}= J^i_T(x(\tau)),
\ee
where $\tau$ is an artificial time with no direct physical interpretation.  This equation enables to compute the streamlines of the transition current. 
\item Drag each point until it reaches the product state \( P\)
\item Start again from \(\partial{A}\), but solve Eq. (\ref{TPTstream})  backward in time, until each point has entered the reactant state \( R \)
\end{enumerate}
Further details about the implementation of the algorithm are given in the Supplementary Material.

\section{The SCPS Algorithm}\label{SCPSalg}
The reader interested in the  derivation of the SCPS approach starting from the unbiased non overdamped Langevin equation is referred to the original pubblication \cite{SCPS}. In this appendix, we only describe the resulting algorithm:
\begin{enumerate}
\item A set of $N_R$ different initial conditions in the reactant state $\{ x_0^{(1)}, \ldots, x_0^{(N_R)}\}$ is generated by sampling some distribution $\rho_0(x)$.  For example, in SCPS applications to protein folding simulations the initial configurations may be generated by running high-temperature MD simulations starting from the crystal native structure. 
\item  From each of these $N_R$ initial conditions, an ensemble of $N_T$ reactive trajectories is calculated using a rMD dynamics based on some predefined reaction coordinate $z(x)$, which 
measures a square distance from some target state $x_0$, i.e.  $z(x) = ||x- x_{0}||^2$. 
In applications of rMD to  complex molecular transitions,  the specific choice of this norm crucially affects the computational efficiency. 
For example,  in rMD  simulations  of protein folding,   the definition  of $z$ based on the Root-Mean-Square-Deviation does not enable to efficiently generate productive folding pathways. 
A much higher folding efficiency is obtained using a Frobenious-type norm between the instantaneous and target continuous atomic 
contact maps~\cite{rMD2}:
\be\label{FB}
z(x) \equiv || \hat C(x)- \hat C(x_0)||^2_F = \frac{2}{N (N-1)}  \sum_{|i-j|>35} (C_{ij}(x) - C_{ij}^0)^2
\ee
where $\hat C(x)$ is the contact map matrix with entries given by
$
C_{ij}(x) = (1-\left(\frac{r_{ij}}{\bar r_0}\right)^6)/(1-\left(\frac{r_{ij}}{\bar r_0}\right)^{10})$ with $r_0= 7.5$\AA.
The constraint $|i-j|>35$ in Eq. (\ref{FB}) is introduced in order to avoid  a bias force between  atoms which are within the same amino-acid.

Standard MD is turned into a rMD by introducing  into the equations of motion the following unphysical history-dependent biasing force:
\begin{equation}\label{rmd_force}
\textbf{F}_{\text{rMD}} = -\frac{k_R}{2}\nabla z(\textbf{x}) \cdot \xi( z(\textbf{x}) - z_{\text{min}}(t))\theta( z(\textbf{x})-z_\text{min}(t))  
\end{equation}
where $z_\text{min}(t)$ denotes the smallest value assumed by $z$ up to time $t$ and $k_R$ is an elastic constant and $\xi(x)$ is a function which is positive definite for $x>0$ (in the present work we used $\xi(x)= x)$. We note that the Heaviside step function switches off the biasing force any time the system has progressed towards the target, according to collective variable $z$ (i.e. for $z(t)< z_\text{min}(t-\Delta t)$).

\item  $N_R$ mean paths are calculated by averaging the $N_T$ trajectories in each of the $N_R$ ensembles. For example, suppose $\{x^{(k)}_{1}(\tau), \ldots, x^{(k)}_{N_T}(\tau)\}$ are the $N_T$ trajectories in feature space generated by rMD starting from the $k-$th initial condition $x_0^{(k)}$. Then we compute:
\be\label{avF}
\langle x(\tau) \rangle_{(k)} = \frac{1}{N_T}\sum_{l=1}^{N_T} x^{(k)}_l(\tau) 
\ee
for all times $\tau$ in the simulation time interval, $\tau\in [0,t_f]$. For protein folding simulations, the average path is computed in contact map space, $\langle f_{ij}(\tau) \rangle_{k} =  \frac{1}{N_T}\sum_{l=1}^{N_T} f^{(k)}_{ij~l}(\tau) $.

\item  The $N_R$ average paths obtained in the previous step are used to define the following  two collective coordinates:
\be\label{s_lambda}
s_\lambda({\bf x})  &=& \frac{1}{t_f}\frac{\int_0^{t_f} dt' t' e^{-\lambda~|| x(\tau) - \langle x(t') \rangle_{(k)} ||^2}}{\int_0^t dt' e^{-\lambda~||x(\tau) - \langle x(t') \rangle_{(k)} ||^2}}.\\
\label{w_lambda}
 w_\lambda(\textbf{x}) &=& -\log \int_0^{t_f} dt' e^{-\lambda~|| x(\tau) - \langle x(t') \rangle_{(k)} ||^2}, \qquad (k= 1, \ldots, N_R). 
 \ee
where $\lambda$ is some arbitrary parameter which must be  chosen  $\gg 1$. In  applications to protein transitions the Frobenious norm (\ref{FB}) is adopted. 

Eqs. (\ref{s_lambda}) and (\ref{w_lambda}) introduced in Ref. ~\cite{tubevar}. 
Their geometrical interpretation is most evident after discretizing the time integrations and is illustrated by Fig. \ref{sigmaFig}. In the  $\lambda\to \infty$ limit, the mathematical function $s_\lambda(x)$ associates to any given   configuration $x$  one specific time $ \bar{t}\in [0,t_f]$ in the average path, namely the one which minimizes the distance  $|| \langle  x(t)) \rangle^{k}- x||^2$:
\be
\min_{\tau\in[0,t_f]} || \langle  x(\tau)\rangle^{k}-  x||^2.
\ee
 In other words, this procedure identifies a projection of the configuration $x$ onto the  the average path. 
 The time $\bar{t} = t(x)$ is then normalized  with respect to the total simulation time $t_f$. As a result, $s_\lambda(x)\in [0,1]$ measures the progress of the reaction, using as reference the  average path.
 Similarly, $w_\lambda(x)$ can be shown to measure the shortest distance between the configuration and the average path.  

\item The collective variables calculated in the previous step are used to define a rMD dynamics with a  self-consistently  defined biasing force given by
\be
\label{scps_force}
{\bf F}_{SCPS}({\bf x}, \tau) &=& -k_w \nabla w_\lambda \cdot (w_\lambda({\bf x}) - w_\text{min}(\tau)) \theta(w_\lambda({\bf x}) - w_\text{min})\nn
&&  + k_s \nabla s_\lambda \cdot (s_\text{max}(\tau)-s_\lambda({\bf x})  ) \theta(s_\text{max}(\tau) - s_\lambda({\bf x})  ).
\ee
 Here,  $w_\text{min}$ ($s_\text{max}$) represents the smallest (largest) value assumed by $w_\lambda$  ($\sigma_\lambda$) up to time $\tau$.  We note that the first biasing force controls how far the trial rMD path can travel from the reference path,  while the second force discourages backtracking towards the reactant.  
\item For each of the $N_R$ initial  conditions, step 3 through 6 are repeated until convergence.  In practice, a SCPS may be considered converged if the
$\mathcal{L}_2$ distance between  the average paths   calculated at different iterations falls below a given threshold $\epsilon$:
\be\label{convergence}
\int_0^t d\tau (\langle x(\tau) \rangle_{n+1}-\langle x(\tau) \rangle_{n})^2< \epsilon
\ee
where $n$ labels the SCPS iteration. 
In addition, in the specific case of protein folding, convergence may be further assessed using the scheme developed in Ref. \cite{BFA} based on the so-called path similarity parameter introduced in Ref. \cite{rMD2}.
\end{enumerate}
An enhanced path sampling approach based on the self-consistent definition of the collective variables (\ref{s_lambda}) and (\ref{w_lambda}) was first proposed in Ref. \cite{PATHmeta}. However, the interfacing with rMD makes the present approach applicable to investigating folding transitions of large proteins. For example, in Ref. \cite{BJ} a variationally improved version of rMD called Bias Functional approach was used to characterize folding and misfolding of proteins consisting of almost 400 aminoacids and folding on the time scales of tens of minutes. 

\section{Relationship between iso-$\sigma$ and iso-$q^+$ curves}
\label{q_param_sigma}
To address this point, we first consider the simplest case of a reaction  in a smooth energy landscape, in the small temperature limit. In this limit,  the average path calculated by SCPS (lasting for a sufficiently long time) is expected to travel along the minimum energy path (MEP). Thus, at least in the vicinity of this curve ( where the reactive current is largest), we expect $\nabla \sigma(x)$ to be proportional to $\nabla q(x)$, which implies  that $\sigma(x)$ is constant over the iso-committor hyperplanes locally orthogonal to the MEP (see Fig. \ref{sigmaFig} (b)). 

Let us now consider the more general case of diffusion in a arbitrarily rugged energy landscape, at finite temperature.
Using the path integral representation of the transition current (\ref{J_SCR})  (see e.g. discussion in Ref. \cite{Mazzola}) we arrive to the following expression for the transition current, derived in Eq. \be
J^i_{SCR}(x) = \frac{1}{t-t_0} \int_{t_0}^t d\tau \frac{1}{2} \left(\langle v^i(x, \tau) \rangle^{(P)} +\langle v^i(x, \tau) \rangle^{(R)} \right)~Q^{(R)}(x,t-\tau)~P^{(P)}(x,\tau).
\ee
In this equation $ \langle v^i(x, \tau) \rangle^{(P)}$ and $ \langle v^i(x, \tau) \rangle^{(R)}$ respectively denote  the average velocity of the paths at point $x$ at time $\tau$ obtained from the path integral representation of $Q^{(R)}(x,\tau)$ and $P^{(P)}(x,\tau)$:
\be\label{velocities}
\langle v^i(x, \tau) \rangle^{(P)} &=& \frac{\int dx_i \rho_0(x_i) \int_{x_i}^x \mathcal{D}x ~\dot x(\tau) ~e^{-  \int_0^\tau dt'  ( L_{OM}[x] + \Omega_P[x])} }{\int dx_i \rho_0(x_i) \int_{x_i}^x \mathcal{D}x ~ e^{-\int_0^\tau dt' (L_{OM}[x] + \Omega_P[x])} }\\
\langle v^i(x, \tau) \rangle^{(R)} &=& \frac{\int dx_f h_P(x_f) \int_{x}^{x_f} \mathcal{D}x ~\dot x(\tau)~ e^{- \int_\tau^{t} dt' (L_{OM}[x]+\Omega_R[x])} }{\int dx_f h_P(x_f) \int_{x}^{x_f} \mathcal{D}x~ e^{-\int_\tau^{t}dt' (L_{OM}[x] + \Omega_R[x])}},
\ee
where 
$
L_{OM}[x] =  \frac{1}{4D} \left( \dot x + \beta D \nabla U(x) \right)^2
$
is the integrand of the so-called Onsager-Machlup functional and the functions $\Omega_{R}(x)$ and $ \Omega_{P}(x)$ are respectively defined to vanish outside the reactant and product  and to be effectively infinite inside. Thus,  the functionals $\int_0^t dt' \Omega_{P}[x(t')]$ an $\int_0^t d t' \Omega_{R}[x(t')]$ impose absorbing boundary conditions at $\partial_P$ and $\partial_R$. 

As usual, if the time interval are chosen in the SCR, then the  dependence on $\tau$ is suppressed and one arrives to a simple equation:
\be
J^{i}_{SCR}(x) \simeq \langle v^i(x)\rangle~ m_{SCR}(x) 
\ee
where $\langle v^i(x) \rangle \equiv \frac{1}{2} \left(\langle v^i(x) \rangle^{(P)} +\langle v^i(x)\rangle^{(R)}\right)$ is the average velocity of the transition paths passing through $x$. 

Let us now choose $x$ to be a point on the principal curve calculated using the SCPS algorithm. Then,  the transition current $J^i_{SCR}$ is parallel to the time derivative of the average position $\langle x(\tau)\rangle$ entering the definition of the collective variable $\sigma(x)$ ---cfr. Eq. (\ref{sigma})---. Since the transition current is proportional to $\nabla q^+$,  in the vicinity of the principal curve iso-committor and iso-$\sigma$ hyperplanes coincide. 

\section{Derivation of the Conditional Langevin Dynamics Equation in the SCR}
\label{CLD_app}
The first step in our derivation consists in recalling that, in the  SCR regime,  the  ergodic transition path density distribution $m_T(x)$  can be rigorously approximated with the   $m_{SCR}(x)$  distribution -- defined in Eq. (\ref{m_SCRdef}) ---  which is sampled by short, non-ergodic reactive pathways.  
 Using Eq. (\ref{m_SCR}) and  assuming that the time scale $t_f$ is in the SCR,  $m_{SCR}(x) $ can be conveniently re-written in the following form:
\be\label{mSCRQP}
m_{SCR}(x,t) = \frac{1}{t_f-\tau_0} \int_{\tau_0}^{t_f} dt ~\mathcal{P}(x,t),
\ee
where
\be\label{PQP}
\mathcal{P}(x,t)=  \frac{1}{\mathcal{N}}~Q^{(R)}(x,t)P^{(P)}(x,t)
\ee
 and $\mathcal{N}$ is a normalisation factor which does not need to be specified.

It is straightforward to show that $\mathcal{P}(x, t)$ obeys the following partial differential equation \cite{CLD1} 
\be\label{PDE}
\frac{\partial}{\partial t} \mathcal{P}(x,t) &=&  D\nabla\cdot \left( \nabla + \beta \nabla U(x) - 2 \nabla \ln Q^{(R)}(x,t)  \right) \mathcal{P}(x,t) 
\ee
We emphasize that the microscopic dynamics underlying Eq. (\ref{PDE}) is the same of the original Langevin equation. This  is reflected by the fact that Eq. (\ref{PDE}) has the structure of a SFP equation with the additional  term $- 2 \nabla \ln Q^{(R)}(x,t)$,  which arises from imposing specific boundary conditions.  

This implies that the   $\mathcal{P}(x,t)$ distribution can be sampled by integrating an effective Langevin equation with a history-dependent biasing force:
\be\label{LB}
\dot x &=&  D  \left(-  \beta \nabla U (x) + 2  \nabla \ln Q^{(R)}(x,t) \right) + \eta(t) 
\ee

Using the results concerning the spectral representation of the $Q^{(R)}(x,t)$ Green's function derived in the previous section --Eq.s  (\ref{QRapp}) through (\ref{q^+_SCR}) -- we obtain an explicit and time-independent expression for the biasing force, in terms of the SCR estimate of the committor:
\be\label{LB3_app}
\dot x &=&  D  \left(-  \beta \nabla U (x) +  2 \beta^{-1} \frac{\nabla q_{SCR}^+(x)}{q_{SCR}^+(x)} \right) + \eta(t) 
\ee
thus recovering Eq. (\ref{LB2}).  

\section{Sampling the Bolzmann Distribution by irMD }
\label{proveirMD}
In this appendix we show  that the irMD introduced in section \ref{irMDalg}  can be used to sample the Boltzmann distribution restricted to the reactive region. 
To this goal we begin by  choosing $\xi(x)\equiv 1$ in Eq. (\ref{irMD}). 

Let us now consider the same  stochastic process adopted to sample $m_T(x)$ using  CLD simulations ---cfr. section \ref{CLDalg}---: We generate many reactive trajectories by integrating the irMD, initiating a new one in the reactant,  anytime a reactive trajectory reaches the product. 

In the reactive region and at a sufficiently large distances from the borders of the reactant and product states  the probability density generated by this irMD  obeys the  following master equation:
\be\label{master}
\partial_t P(x,t) = \int dy' \left(P(y', t) \tilde{W}(y|y') - P(y,t) \tilde{W}(y'|y)\right)
\ee
In this equation, $\tilde{W}(y'|y)$ is transition probability from $y$ to $y'$ per unit time  $\Delta t$ in the  irMD dynamics and can be written as follows
\be
\tilde{W}(y'|y) = W_0(y'|y) \theta(q^+(y')-q^+(y)) + W_B(y'|y) \theta(q^+(y)-q^+(y')), 
\ee
where $W_0(y'|y)$ is the transition rate in the original (i.e. unbiased) Langevin dynamics,  while $W_B(y'|y)$ is the corresponding rate  the biased Langevin dynamics, i.e.  with the additional  force $F_{B}(x) = k_R \nabla q(x)$. 
We emphasize that for sake of simplicity we have chosen to consider the evolution of  probability distribution   in the bulk of the transition region, i.e. where it is possible to neglect the contribution of the transitions from the reactant to the point $x$ and from point $x$ to the product. A generalized discussion which  includes such transitions within the framework of discrete jump Markov processes  is reported in Ref.s~\cite{CLDMSM1, CLDMSM2}. 

The Fokker-Planck equation associated to the master equation (\ref{master}) can be obtained by explicitly computing the first two Kramers-Moyal coefficients (here we discuss the 1D case only,  for sake of notational simplicity)
\be
a_1(x) &\equiv& \lim_{\Delta t\to 0} \frac{1}{\Delta t} \left[\int dy' (y - y) \tilde{W}(y'|y) \right]\nn
\label{a1}
&&\hspace{-1.5cm}=  \lim_{\stackrel{\Delta t\to 0}{x(t)\to x}} \frac{1}{\Delta t} \Big[ \langle \big( x(t+\Delta t)- x(t) \big) \theta \big[ q^+[x(t+\Delta t)] - q[x(t)]  \big] \rangle_0 + \langle \big( x(t+\Delta t)- x(t)\big) \theta\big[ q^+[x(t)] - q^+[x(t+\Delta t] \big]  \rangle_B \Big] \\
a_2 (x) &\equiv& \lim_{\Delta t\to 0} \frac{1}{\Delta t} \left[\int dy' (y' - y)^2 \tilde {W}(y'|y) \right] \nn
\label{a2}
&&\hspace{-1.5cm}=  \lim_{\stackrel{\Delta t\to 0}{x(t)\to x}} \frac{1}{\Delta t} \Big[ \langle \big( x(t+\Delta t)- x(t) \big)^2 \theta \big[ q^+[x(t+\Delta t)] - q[x(t)]  \big] \rangle_0 + \langle \big( x(t+\Delta t)- x(t)\big)^2 \theta\big[ q^+[x(t)] - q^+[x(t+\Delta t] \big]  \rangle_B \Big], \qquad
\ee
 $\langle \ldots \rangle_B$ and $\langle \ldots \rangle_0$ denote respectively an average of over the Gaussian noise  in the Langevin dynamics with and without the biasing force. 
 The averages in Eq.s (\ref{a1}) and (\ref{a2}) can be calculated by explicitly performing the  integral over the  Gaussian noise, relying on the integral representation of the Heaviside step function. We find
   \be
a_1(x) &=& -\frac{D}{k_BT}\left( \nabla U(x) - \frac{k_R}{2} \nabla q(x)\right) \\
a_2(x) &=& 2 D.
\ee
Therefore, the  Fokker Planck associated to the master equation (\ref{master}) is simply
\be
\partial_t P(x,t) = D \nabla \left(\nabla + \beta \nabla U(x) - \frac{k_R}{2} \nabla q(x) \right) P(x,t)
\ee
We emphasize that this equation holds for any value of the coupling constant $k_R$.  It is straightforward to verify that the Boltzmann distribution provides a stationary solution of this equation, owing to the fact that the committor obeys the backward Kolmogorov equation. 

We stress the fact that this stationary solution does not correspond to an equilibrium state, but to the stationary  distribution of a nonequilibrium process with emitting and absorbing boundary conditions.  As a result, the corresponding Fokker-Planck probability current is proportional to the transition current of TPT:
\be
J_{irMD}(x) = -D\left(\nabla + \beta \nabla U(x) - \frac{k_R}{2} \nabla q(x) \right) \frac{e^{-\beta U(x)}}{Z} =  \frac{k_R}{2} \nabla q(x) \frac{e^{-\beta U(x))}}{Z}. 
\ee

Let us now consider the strong bias limit, $k_R\to \infty$. In this regime,  the choice of the function $\xi(x)$ entering the definition (\ref{irMD}) of the irMD biasing force  becomes irrelevant, because the backtracking parts of the reactive trajectories  have negligible extent.  Therefore, for sufficiently large values of $k_R$, we expect the irMD dynamics to provide the exact sampling of the Boltzmann distribution restricted to the reactive region, regardless of the choice of $\xi(x)$.

\newpage

\section*{Supplementary Material}

\section*{S1. Calculation of the Average Path in SCPS Iterations}
\label{app:cut}

The mean path was computed by averaging over time the rMD trajectories which reached the product state. First, we identified the reactant and the product  as the regions around \(x_R =(-1,0) \) and \(x_P = (1,0)\) respectively, where the potential doesn't exceed the threshold \(V_t = -2.5 k_BT\). This choice is used to impose the boundary conditions for the committor function: \( q(x) \big \rvert_R = 0  \) and \( q(x) \big \rvert_P = 1  \). A trajectory was assumed to have arrived to the product state only if the minimal value of the biasing coordinate (the Euclidean distance from the target) reached a value lower than $z_{t} = 0.02$. 
The only portion of the trajectories which is relevant for the definition of mean path is the reactive one. For this reason it is desirable to cut the trajectories as soon as they cross the boundaries, discarding all frames corresponding to fluctuations in these basin regions. On the other hand, the definition of mean path $\langle f(t)\rangle$ involves averaging over trajectories which have identical number of time frames (cfr. Eq. (B3) in appendix B of the main text).  Therefore, we adopted the following prescription to select the fixed duration of the time window \(t_f \) entering Eq. (B3). 
We started storing the times \( t_P\) and \( t_R \)  at which each trajectory entered the product and exited the reactant, respectively. Subtracting them we collected 
$ t_{\text{react}}= t_P-t_R $, the time spent in the reactive region by each trajectory. We found a skewed distribution, as expected from general mean first passage time considerations. 
We computed the mean \( \tau \) and the standard deviation \(\Delta \tau \) of this distribution, and set the time interval $t_f$  equal to
$
t_f = \tau + \Delta \tau.
$

Then, we selected out equally long segments from each trajectory, by harvesting the frames between the times \( t_1 = t_P - \frac{t_f - t_{ \text{react} }}{2}\) and \( t_2 = t_R + \frac{t_f-  t_{ \text{react} }}{2} \).  Finally, we discarded the trajectories that within $t_f$  didn't connect the reactant and the product states.

The frequency histogram of configurations visited by trajectories generated with SCPS during the first (a), second (b) and third (c) iteration for the two dimensional toy model discussed in the
main text are reported in Fig. \ref{fig:SCPS_examples}. 
\begin{figure}[t!]
\centering
\includegraphics[width=\textwidth]{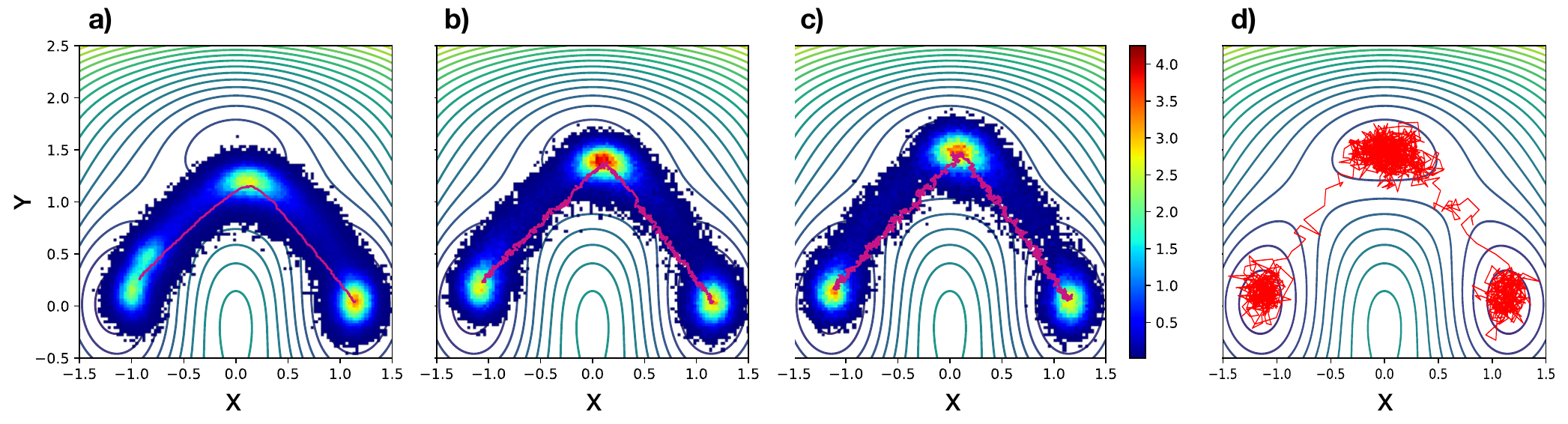}
\caption{Frequency histogram of configurations visited by trajectories generated with SCPS during the first (a), second (b) and third (c) iteration. The superimposed purple curve represents the average path of the corresponding iteration. (d) Typical trajectory obtained from the third iteration of SCPS. The background of all the four pictures represents a contour plot of the potential.}{\label{fig:SCPS_examples}}
\end{figure}

\section*{ S2. Choice of the parameter $\lambda$}
\label{app:lambda}

\begin{figure}[t!]
\center 
\includegraphics[width=\textwidth]{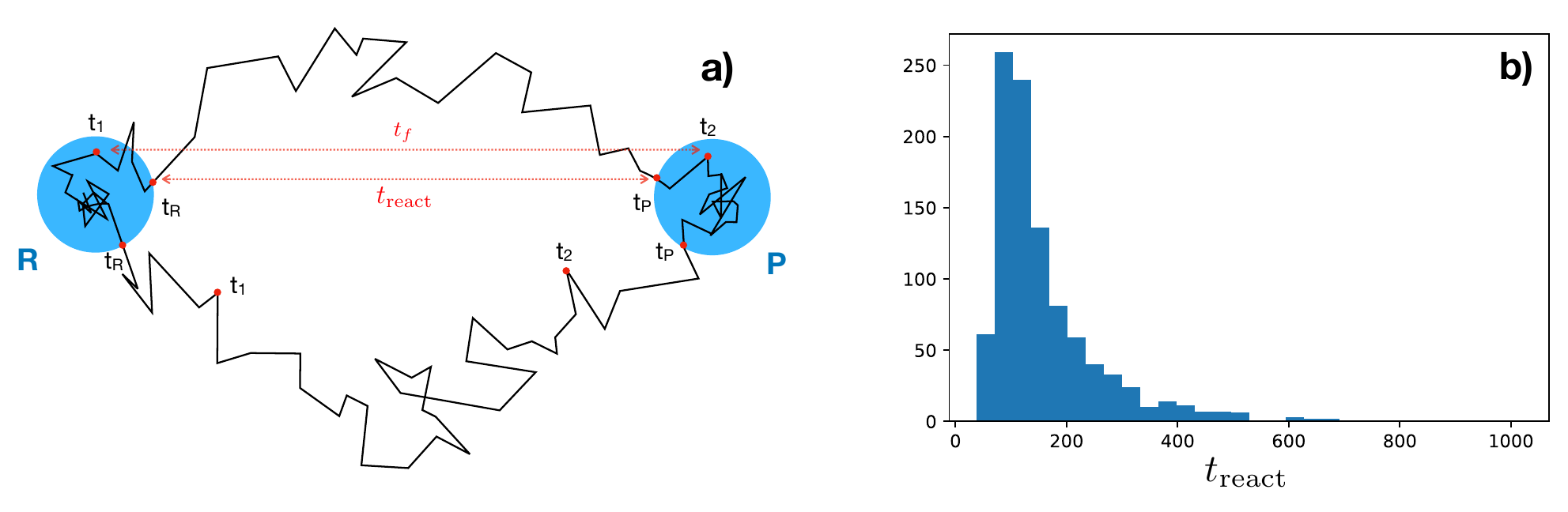}
\caption{a) Example of two trajectories which contain a reactive part. The upper one is retained in the cut procedure, since \(t_f >  t_{\text{react}} \), while the second one is not. b) Example of reactive times histogram, in time step units, related to the rMD stage }
\label{fig:cut_drawing}
\end{figure}

The collective coordinate $\sigma(x)$ defined in Eq. (29) of the main text involves a free parameter $\lambda$ which must be chosen in such a way to satisfy the condition $\lambda\gg 1$. Indeed, only asymptotically large values of $\lambda$ the collective coordinate $\sigma(x)$ selects out  the  frame in the average path $\langle {\bf x}(t)\rangle$  which is nearest to the configuration $x$ (see discussion in the main text).  In particular, our simulations were performed chosing \(\lambda = 30\), and retaining only \(N=100\) points of the mean path. To ensure that this choices are consistent with the asymptotic limit  condition we picked a few values of $x$ in the reactive region and examined how many frames in the average path $\langle {\bf x}(\tau)\rangle$  had contributed significantly to the time integral defining $\sigma$.  Fig. \ref{fig:reactive_time}  shows that, indeed,  with our choice of $\lambda$ only a few frames influence the computation of $\sigma$.

\section*{ S3. Calculation of the committor from unbiased Langevin dynamics}\label{app:true_committor}
In order to obtain an reference calculation of the committor $q^+(x)$ associated to the potential in Eq. (36) of the main text we adopted the following procedure. The region $[-1.5, 1.5] \times [-0.5, 2.5]$ was partitioned in $100\times 100$ bins. From each block we randomly sampled $2\times 10^5$ points which we used as initial conditions for just as many Langevin dynamics simulations. In particular, Langevin dynamics simulations consisting of $4 \times 10^3$ time steps were performed using $dt=0.02$, $\gamma=1$ and $1/\beta = 0.15$. A trajectory was considered to have reached the product if, at some point $(x,y)$ along the simulation, $U(x,y) < -2.5$ k$_B$T and $z=\sqrt{(x-x_t)^2+(y-y_t)^2} < 0.02$, where $(x_t, y_t)$ was the target point of the biased simulation. If this condition was met, the trajectory was stopped. The value of the committor in the bin $(i,j)$ was then computed as
\begin{equation}
q^+(i,j) = \frac{\text{Number of simulations started from } (i,j) \text{ which reached the product before the reactant}}{\text{Total number of simulations started from } (i,j)}
\end{equation}

\begin{figure}[t!]
\center 
\includegraphics[width=\textwidth]{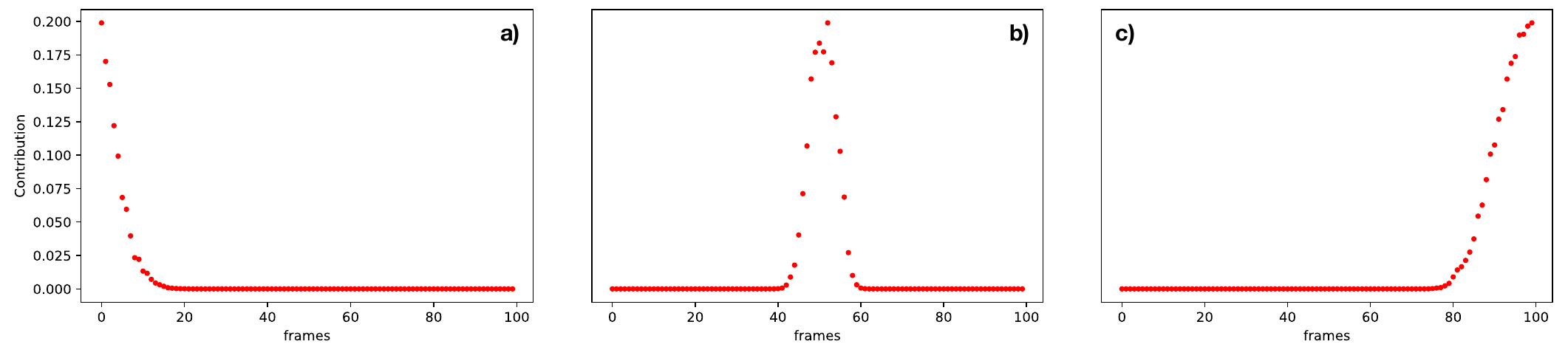}
\caption{Relative contribution to the collective variable $\sigma(x)$ of the different frames in the average path $\langle {\bf x}(\tau) \rangle$. These results  were calculated for three representative positions, near the beginning the center and the end of the reactive region. }
\label{fig:reactive_time}
\end{figure}

\section*{ S4. Calculation of the committor from SCPS simulations}\label{app:minimization}
\begin{figure}[!h]
\includegraphics[scale=0.7]{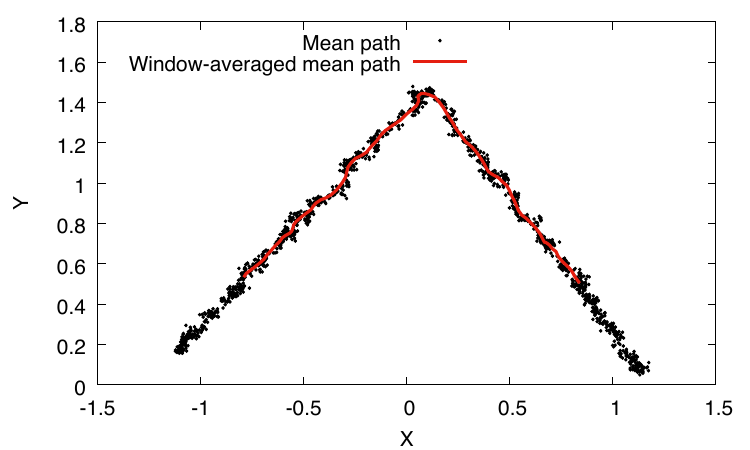}
\caption{Result of window-averaging on the mean path obtained from the last SCPS iteration, after all the points satisfying $U(x,y) <  -2.5$ k$_B$T were removed. }
\label{fig:S1}
\end{figure}
In order to compute the committor $q^+(x)$ from the results of the SCPS simulations we adopted the following procedure. First, the points in the path such that $U(x,y) < -2.5$ k$_B$T were discarded because they were considered as belonging to the product or reactant basins. Then the so-obtained path was window averaged using a window of 50 points: the result is showed in Fig.S1. After this procedure, we retained only $N=100$ points for the window-averaged path. 

The result of this procedure was applied as the initial condition for the functional optimization strategy discussed in section III  of the main text. In order to minimize the functional in Eq. (31) of the main text, we employed a Sequential Least SQuares Programming (SLSQP) algorithm \cite{minimization1, minimization2}, constraining the initial and the final points to be respectively $q(0)=0$ and $q(N)=1$. The initial guess for the values of $q^+(\sigma(x))$ was provided by the values of $\sigma(x)$ computed for the points along the window-averaged mean path. The calculation converged after 101 iterations using a precision goal for the value of the functional in the stopping criterion of $10^{-8}$. 

\section*{Calculation of the reactive current streamlines and flux tubes}
\label{app:tubes}

The reactive current provides informations complementary to the reactive probability density. These can be displayed plotting the reactive flux tubes, that we constructed according to the procedure explained in appendix A of the main text. Restricting to iso-committor curves, the normal to the area element is parallel to \( \nabla q^+\) so the flux flowing through such curves is simply the integral of the equilibrium probability density
\be
\label{flux}
\sum_{i=1}^{3N} \int_{\partial{S}}  J^i_T(x) dS_i(x) = \sum_{i=1}^{3N}\int_{\partial{S}} dx \; e^{-\beta U(x)} \; \left( \nabla_i q^+(x) \hat{n}_i(x) \right) = N \int_{\partial{S}} dx \; e^{-\beta U(x)}
\ee 
In particular, we selected points belonging to the iso-committor surface \(\partial{S}\) defined by \( S \equiv \left \{ x \; | \; q^+_{SCR}(x)=0.5 \right \} \), we have integrated the equilibrium probability distribution restricted to this iso-line and located the peak of this restricted distribution. Then we have defined the window \(\partial{A}\) as the portion of \(\partial{S}\) centered at this peak that encloses the \(30\%\) of the probability, i.e.
\be
\int_{\partial{A}} dx \; e^{-\beta U(x)} =  0.3\; \int_{\partial{S}} dx \; e^{-\beta U(x)}
\ee 
The endpoints of this window were used as starting points for the artificial evolution regulated by  Eq. (A6) in appendix A of the main text. We integrated this differential equation forward in the artificial time \( \tau \) until we reached the product state \(P\) and backward until we reached the reactant state \(R\).  
We repeated this procedure for the fraction of \(60\%\) and the \(90\%\) of the probability. The results are shown in Fig. 7 of the main text.

\end{document}